\documentclass[twocolumn,showpacs,amsmath,amssymb]{revtex4}

\usepackage{epsfig,xspace}
\usepackage{graphicx}
\usepackage{dcolumn}
\usepackage{bm}

\newcommand{\cuteval}{{\sc CutEval}\xspace}

\begin{document}

\title{ Observation of High Energy Atmospheric Neutrinos
 with the Antarctic Muon and Neutrino Detector Array }

\author{
J.~Ahrens,$^{9}$
E.~Andr\'es,$^{14}$ 
X.~Bai,$^{1}$
G.~Barouch,$^{11}$
S.W.~Barwick,$^{8}$ 
R.C.~Bay,$^{7}$ 
T.~Becka,$^{9}$ 
K.-H.~Becker,$^{2}$ 
D.~Bertrand,$^{3}$ 
F.~Binon,$^{3}$
% ...
\\
\vspace*{1ex}
(AMANDA Collaboration)
\vspace*{1ex}
}
\affiliation{$^1$ Bartol Research Institute, University of Delaware, Newark, DE 19716, USA
}
\affiliation{
$^2$ Fachbereich 8 Physik, BUGH Wuppertal, D-42097 Wuppertal, Germany 
}
\affiliation{
$^3$ Universit\'e Libre de Bruxelles, Science Faculty CP230, Boulevard du Triomphe, B-1050 Brussels, Belgium 
}
\affiliation{
$^4$ DESY-Zeuthen, D-15735 Zeuthen, Germany 
}
\affiliation{
$^5$ Dept. of Technology, Kalmar University, S-39182 Kalmar, Sweden   
}
\affiliation{
$^6$ Lawrence Berkeley National Laboratory, Berkeley, CA 94720, USA   
}
\affiliation{
$^7$ Dept. of Physics, University of California, Berkeley, CA 94720, USA   
 }
\affiliation{
$^8$ Dept. of Physics and Astronomy, University of California, Irvine, CA 92697, USA   
}
\affiliation{
$^9$ Institute of Physics, University of Mainz, Staudinger Weg 7, D-55099 Mainz, Germany   
}
\affiliation{
$^{10}$ Dept. of Physics and Astronomy, University of Pennsylvania, Philadelphia, PA 19104, USA  
}
\affiliation{
$^{11}$ Dept. of Physics, University of Wisconsin, Madison, WI 53706, USA    
 }
\affiliation{
$^{12}$ Physics Department, University of Wisconsin, River Falls, WI 54022, USA    
}
\affiliation{
$^{13}$ Division of High Energy Physics, Uppsala University, S-75121 Uppsala, Sweden     
}
\affiliation{
$^{14}$ Dept. of Physics, Stockholm University, SCFAB, SE-10691 Stockholm, Sweden    
}
\affiliation{
$^{15}$ Vrije Universiteit Brussel, Dienst ELEM, B-1050 Brussel, Belgium  
}

\date{\today}

\begin{abstract}
The Antarctic Muon and Neutrino Detector Array (AMANDA) 
began collecting data
with ten strings in 1997.  Results from the first year of operation are
presented.  Neutrinos coming through the Earth from the Northern Hemisphere
are identified by secondary muons moving upward through the array.  
Cosmic rays in the
atmosphere generate a background of downward moving muons, which are about
$10^6$ times more abundant than the upward moving muons.  
Over 130 days of 
exposure, we
observed a total of about 300 neutrino events.
In the same period, a
background of $1.05\cdot 10^9$ cosmic ray muon events was recorded.  The
observed neutrino flux is consistent with atmospheric neutrino predictions. 
Monte Carlo simulations indicate that 90\% of these events lie in the
energy range $66$\, GeV to $3.4$\,TeV. The observation of atmospheric
neutrinos consistent with expectations establishes AMANDA-B10 as a working
neutrino telescope.
\end{abstract}


\maketitle


\section{Introduction}

Energetic cosmic ray particles entering the Earth's atmosphere generate a
steady flux of secondary particles such as electrons, muons and neutrinos. 
The electronic component of cosmic rays is quickly absorbed.
High energy muons penetrate the Earth's surface
for several kilometers, while atmospheric neutrinos can easily
pass the Earth up to very high energies.  
Interactions of hadronic particles, similar to the ones that
create the atmospheric neutrino flux, will generate neutrinos at 
sites where cosmic rays are generated and where they interact 
as they travel through the 
Universe.  The goal of observing 
neutrinos of astrophysical origin determines the design
and the size of neutrino telescopes.

The primary channel through which neutrino telescopes detect neutrinos above 
energies of a few tens of GeV is by observing the Cherenkov light from 
secondary muons produced in \mbox{$\nu_\mu$-nucleon} interactions in or near 
the telescope. To ensure that the observed muons are produced by neutrinos, the
Earth is used as a filter and only upward moving muons are selected. 
A neutrino telescope consists of an array of photosensors embedded deeply 
in a
transparent medium. The tracks of high energy muons --- which can travel 
many hundreds of meters, or even kilometers, through water or ice --- can be
reconstructed with reasonable precision even with a coarsely instrumented 
detector, provided the medium is sufficiently transparent. A location deep 
below the surface serves to minimize the flux of cosmic-ray muons.

In this paper we
demonstrate the observation of atmospheric muon neutrinos with the Antarctic
Muon and Neutrino Detector Array (AMANDA).  These neutrinos constitute a
convenient flux of fairly well known strength, angular distribution, and
energy spectrum, which can be used to verify the
response of the detector.
The paper will focus on the methods of data analysis and the comparison of
observed data with simulations.  After a brief description of the detector,
the data and the methods of simulation are introduced in Section~\ref{sec:data}
and the general methods of event reconstruction are described in 
Section~\ref{sec:reco}.  Two AMANDA working groups analyzed the data
in parallel.
The methods and results of both analyses are described in
Sections~\ref{sec:desy} and~\ref{sec:uw}.
After a discussion of systematic uncertainties in Section~\ref{sec:syst} we
present the final results and conclusions.

\section{The AMANDA Detector}

\begin{figure}
\centering
\includegraphics[width=\linewidth]{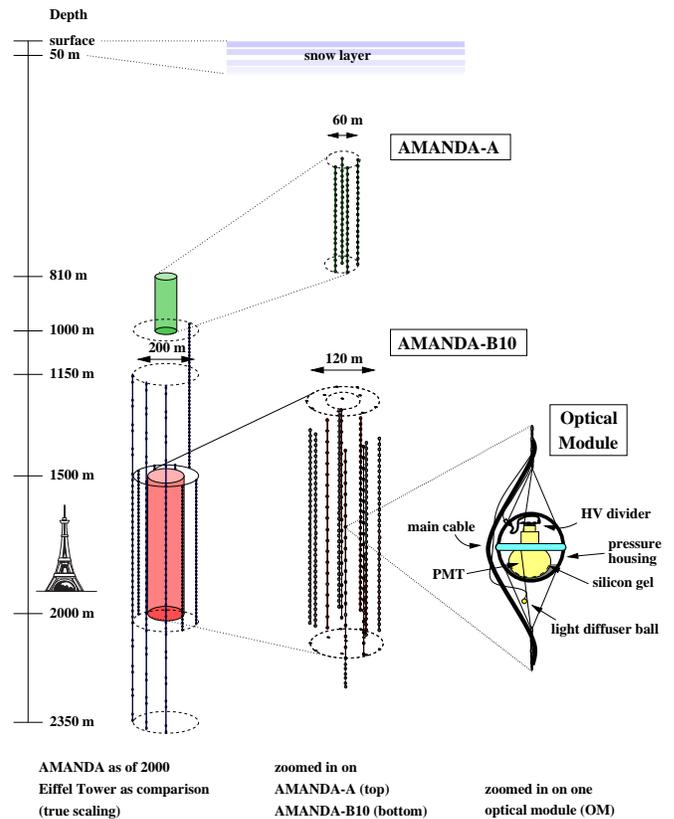}
\caption{The present AMANDA detector.  This paper describes data taken
with the ten inner strings shown in expanded view in the bottom center.}
\label{fig:eiffel}
\end{figure}

The AMANDA detector uses the 2.8\,km
thick ice sheet at the South Pole as a neutrino target, Cherenkov medium
and cosmic ray flux attenuator.  The detector consists of vertical strings of
optical modules (OMs) --- photomultiplier tubes sealed in glass pressure
vessels --- frozen into the ice at depths of 1500--2000 m below the
surface.  Figure~\ref{fig:eiffel} shows the current configuration of the
AMANDA detector.  The shallow array, AMANDA-A, was deployed at depths of
800 to 1000\,m in 1993--94 in an exploratory phase of the project.  Studies
of the optical properties of the ice carried out with AMANDA-A showed a
high concentration of air bubbles at these depths, leading to strong
scattering of light and making accurate track reconstruction impossible. 
Therefore, a deeper array of ten strings with 302 OMs was deployed in the
austral summers of 1995--96 and 1996--97 at depths of 1500--2000\,m.  This
detector is referred to as AMANDA-B10, and is shown in the center of
Fig.~\ref{fig:eiffel}.  The detector was augmented by three additional
strings in 1997--98 and six in 1999--2000, forming the AMANDA-II array.

In AMANDA~B10, an optical module consists of a single 8'' Hamamatsu R5912-2
photomultiplier tube (PMT) housed in a glass pressure vessel.  The PMT is
optically coupled to the glass housing by a transparent gel.  Each module
is connected to electronics on the surface by a dedicated electrical cable,
which supplies high voltage and carries the anode signal of the PMT\@.  For
each event, the optical module is read out by a peak-sensing ADC and a TDC
capable of registering up to eight separate pulses.  The overall precision of
measurement of photon arrival times is approximately 5\,ns.  Details of
deployment, electronics and data acquisition, calibration, and the
measurements of geometry, timing resolution, and the optical properties of
the ice can be found in~\cite{B4}.

\begin{figure}
\centering
\includegraphics[width=8.5cm]{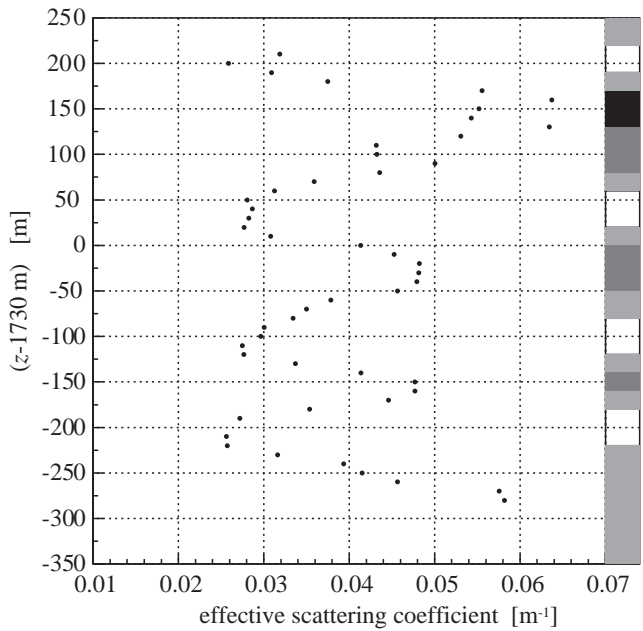}
\caption{
Variation of the optical properties with depth.  The effective
scattering coefficient at a wavelength of 532\,nm is shown as a 
function of depth.  
The $z$ axis is pointing upwards and denotes the vertical distance from
the origin of the detector coordinate system located at a depth of 1730\,m.
The shaded areas on the side indicate layers
of constant scattering coefficient as used in the Monte Carlo.
}
\label{fig:layers}
\end{figure}

The optical properties of the polar ice in which AMANDA is embedded have
been studied in detail, using both light emitters located on the strings and
the downgoing
muon flux itself.  These studies~\cite{kurt_99} have shown that
the ice is not perfectly homogeneous, but rather that 
it can be divided into several
horizontal layers which were laid down by varying climatological 
conditions in the past~\cite{Price}.  Different concentrations of dust 
in these layers lead to a modulation of the scattering and
absorption lengths of light in the ice, as shown in Fig.~\ref{fig:layers}. 
The average absorption length
is about 110\,m at a wavelength of 400\,nm at the depth of the AMANDA-B10 array,
and the average effective scattering length is approximately
20\,m.

\section{Data and Simulation}
\label{sec:data}

The data analyzed in this paper were
recorded during the austral winter of 1997, from April to
November.  Subtracting downtime for detector maintenance, removing runs
in which the detector behaved abnormally
and correcting for deadtime in the data acquisition
system, the effective livetime was 130.1 days.

Triggering was done via a majority logic system, which demanded that 16 or
more OMs report signals within a sliding window of 2\,$\mu$s.  When this
condition was met, a trigger veto was imposed and the entire array read
out.  The raw trigger rate of the array was on average 75\,Hz, 
producing a total
data set of $1.05 \cdot 10^9$ events.

Random noise was observed at a rate of 300\,Hz for OMs on the
inner four strings and 1.5 kHz for tubes on the outer six, the difference
being due to different levels of concentration 
of radioactive potassium in the pressure
vessels (details on noise rates can be found in ref~\cite{sn-bouchta}). 
A typical event has a duration of 4.5 $\mu$s, including the
muon transit time and the light diffusion times, so random noise contributed
on average one PMT signal per event.

\begin{figure}
\centering
\includegraphics[width=8cm]{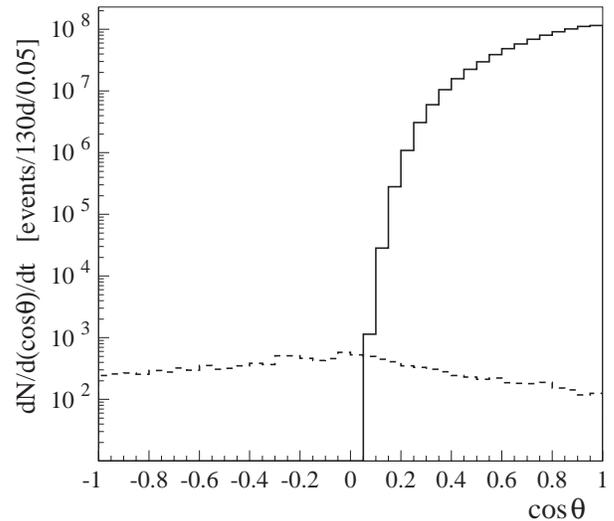}
\caption{The zenith angle distribution of simulated AMANDA triggers per
130.1 days of livetime.  The solid line represents triggers from downgoing
cosmic ray muons generated by {\tt CORSIKA}.  The dashed line shows
triggers produced by atmospheric neutrinos.}
\label{fig:trig_zenith}
\end{figure}

Almost all of the events recorded were produced by downgoing muons
originating in cosmic ray showers.  
Triggers from atmospheric neutrinos contribute only a
few tens of events per day, a rate small compared to the 
event rate from cosmic ray muons, 
as shown in Fig.~\ref{fig:trig_zenith}.  The
main task of AMANDA data analysis is to seperate these neutrino events 
from the background of cosmic-ray muons.  Monte Carlo (MC)
simulations of the
detector response to muons produced by neutrinos or by cosmic rays were
undertaken to develop techniques of background rejection.

Downgoing muons were generated by atmospheric shower simulations of
isotropic protons with {\tt basiev}~\cite{basiev} or protons and heavier
nuclei with {\tt CORSIKA} using the QGSJET generator~\cite{corsika}, and
tracked to the detector with the muon propagation code {\tt mudedx}
\cite{lohmann, lohmann2}.
Two other muon propagation codes were used to check for systematic
differences: {\tt PROPMU}~\cite{lipari} with a 30\% lower rate and {\tt
MMC}~\cite{Chirkin} with a slightly higher rate.  A total of $0.9\cdot10^8$
events were simulated.  Most characteristics of the events generated with
{\tt basiev} were found to be similar to the more accurate {\tt
CORSIKA}-based simulation.  For the latter, the primary cosmic ray flux as
described by Wiebel-Sooth and Biermann~\cite{wiebel} was used.  The
curvature of the Earth has been implemented in {\tt CORSIKA} to correctly
describe the muon flux at large zenith angles.  The event rate based on
this Monte Carlo was 75\,Hz and compares reasonably well with the observed
rate of 100\,Hz (after deadtime correction).  The detector response to
muons was modeled by calculating the photon fields produced by continuous
and stochastic muonic energy losses~\cite{ptd}, and simulating the response
of the hardware to these photons~\cite{amasim}.  Upgoing muons were
generated by a propagation of atmospheric neutrinos, which were tracked
through the Earth and allowed to interact in the ice in or around the
detector or in the bedrock below~\cite{nusim, nusim2}.  Muons that were
generated in the bedrock were propagated using {\tt PROPMU}~\cite{lipari}
until they reached the rock-ice boundary at the depth of 2800\,m.  The
muons were then propagated through the ice in the same way as those from
cosmic ray showers.  The atmospheric neutrino flux was taken from
Lipari~\cite{lipari_flux}.

The Cherenkov photon propagation through the ice was modeled to
create multidimensional tables
 of density and arrival time probability distributions of
the photon flux.  These photon fields were calculated for pure 
muon tracks and for
cascades of charged particles.  A real muon track was modeled as a
superposition of the photon fields of a pure muon track and the stochastic
energy losses based on cascades.  The photon fields were calculated
out to 400\,m from the emission point, taking into account the
orientation of the OM with respect to the muon or cascade.  In the
detector simulation, the ice was modeled as 16 discrete layers, as 
indicated by the
shaded areas in Fig.~\ref{fig:layers}.  The spectral properties of the
photomultiplier sensitivity, the glass, the gel, and, most importantly, the
ice itself were included in the simulation of the photon propagation.  The
probability of photon detection depends on the Fresnel reflectance at all
interfaces, transmittances of various parts, and quantum and collection
efficiencies of the PMT. The relevant physical parameters have been
measured in the laboratory, so that the spectral sensitivity of the OM
could be evaluated.  Two types of OMs, differing in the type of pressure
vessel, were used in the construction of AMANDA-B10.  The inner four
strings (AMANDA-B4) use Billings housings while the outer six strings use
Benthos housings.
(Benthos Inc. and Billings Industries are the
manufacturers of the glass pressure vessels.  Benthos and  Billings 
are registered trademarks of the respective companies.) 
The two types of housing
have different optical properties.  The Benthos OMs have an effective
quantum efficiency of 21\% at a wavelength of 395\,nm for plane-wave
photons incident normal to the PMT photocathode.  Ninety percent 
of the detected
photons are in the spectral range of 345--560\,nm.

An additional sensitivity effect arises from the ice surrounding the OMs.  The
deployment of OMs requires melting and refreezing of columns of
ice, called ``hole ice'' hereafter.  This cycle results in the formation of
bubbles in the vicinity of the modules, which increase scattering and
affects the sensitivity of the optical modules in ways that are not
understood in detail.  Since the total volume of hole ice is small compared
to bulk ice in the detector (columns of 60\,cm diameter, compared to 30\,m
spacing between strings), its effect on optical properties can be treated
as a correction to the OM angular sensitivity.  The increased 
scattering of photons
in the hole ice has been simulated and compared to data taken with laser
measurements {\it in situ} to assess the magnitude of this effect.  This
comparison provides an OM sensitivity correction that
reduces the relative efficiency in the forward direction, but enhances it
in the sideways and backward directions.  The sensitivity in the backward
hemisphere ($90^{\circ}-180^{\circ}$) relative to the 
sensitivity integrated over all angles ($0^\circ-180^\circ$) 
of the optical sensor increases from 20\% to 27\%,
due to this correction, while the average relative sensitivity in the
forward direction ($0^{\circ}-90^{\circ}$) drops from 80\% to 73\%.  In
other words, an OM becomes a somewhat more isotropic sensor.  

The effective angular sensitivity of the OMs was also assessed  using
the flux of downgoing atmospheric muons as a test beam illuminating both
the 295 downward facing OMs and the 7 upward facing OMs.  We assumed that
the response of the upward facing OMs to light from downward muons is
equivalent to the response of the downward facing OMs to light from upward
moving muons.  
Based on this assumption we derived a modified angular
response function (later referred to as {\tt angsens}), which resulted in a
effective reduction of the absolute OM sensitivity in forward direction.
In this model the effective relative sensitivity is 67\% in the forward
hemisphere, 
and 33\% in the backward hemisphere.
This correction will be used to
estimate the effect of systematic uncertainties in the angular response on
the final neutrino analysis.

The simulation of the hardware response included the modeling of 
gains and thresholds and random noise at the levels measured for
each OM. The transit times of the cables and the shapes of the
photomultiplier pulses, ranging from 170 to 360\,ns FWHM, were included in
the trigger simulation.  Multi-photon pulses were simulated as
superimposed single photoelectron waveforms.  In all, some $8\cdot10^5$ 
seconds of cosmic rays were simulated, corresponding to 7\% of the
events contained in the 1997 data set.

\section{Event Reconstruction \label{sec:reco} }

The reconstruction of muon events in AMANDA is done offline, in several
stages.  First, the data are ``cleaned'' by removing unstable PMTs and
spurious PMT signals (or ``hits'') due to electronic or PMT noise.  The
cleaned events are then passed through a fast filtering algorithm, which
reduces the background of downgoing muons by one order of magnitude.  This
reduction allows the application of more sophisticated reconstruction
algorithms to the remaining data set.

Because of the complexity of the task, and in order to increase the robustness of
the results, two separate analyses of the 1997 data set were undertaken. 
Both proceeded along the general lines described above, but differ in the
details of implementation.  
The preliminary stages, which are very similar
in both analyses, are described here.  The particulars of each analysis
will be described 
in Sections \ref{sec:desy} and \ref{sec:uw}.
A more detailed description
of the reconstruction procedure will be published 
elsewhere~\cite{RECO}.

\subsection{Cleaning and Filtering   \label{sec:clean_filter} }

The first step in reconstructing events is to clean and calibrate the data
recorded by the detector.  Unstable channels (OMs) are identified and
removed on a run-to-run basis.  On average, 260 of the 302 OMs deployed are
used in the analyses.  The recorded times of the hits are corrected for
delays in the cables leading from the OMs to the surface electronics and
for the amplitude-dependent time required for a pulse to cross the
discriminator threshold.  Hits are removed from the event if they are
identified as being due to instrumental noise, either by their low
amplitudes or short pulse lengths, or because they are isolated in space
by more than 80\,m and time by more than 500\,ns 
from the other hits recorded in the event.  Pulses with short 
duration, measured as the time over threshold (TOT), are often related to
electronic cross-talk in the signal cables or the surface electronics.  In
Analysis~II, TOT cuts are applied to individual channels beyond the
standard cleaning common to both analyses (see Section \ref{sec:uw}).  

Following the cleaning and the calibration, a ``line fit'' is calculated for
each event.  This fit is a simple $\chi^2$ minimization of the apparent
photon flux direction, for which an analytic solution can be calculated
quickly~\cite{stenger} (see also~\cite{B4}).  
It contains no details of Cherenkov radiation or
propagation of light in the ice.  
Hits arriving at time $t_{\mathrm{i}}$ at PMT $i$  located at 
$\vec{r_i}$
are projected onto a line.
The minimization of  $ \chi^2 = \sum_i 
( \vec{r}_i - \vec{r}_0 - \vec{v}_{\mathrm{lf}} \cdot t_i )^2 $ gives 
a solution for $\vec{r}_0$ and a velocity $\vec{v}_{\mathrm{lf}}$.
The results of this fit -- at the first stage the direction 
$\vec{v}_{\mathrm{lf}}/ |\vec{v}_{\mathrm{lf}}|$, at later stages
the absolute value of the velocity -- are used to
filter the data set.  Approximately 80--90\% of the data, for which the
line fit solution is steeply downgoing, are rejected at this stage.

\subsection{Maximum Likelihood Reconstruction \label{sec:max_likelihood} }

After the data have been passed through the fast filter, tracks are
reconstructed using a maximum likelihood method.  The observed photon
arrival times do not follow a simple Gaussian distribution attributable to
electronic jitter; instead, a tail of delayed photons is observed.  The
photons can be delayed predominantly by scattering in
the ice that causes them to travel on paths longer than the length of the
straight line inclined at the Cherenkov angle to the track.  Also, photons
emitted by scattered secondary electrons generated along the track 
will have emission angles
other than the muon Cherenkov angle.  These effects generate a distribution
of arrival times with a long tail of delayed photons.

\begin{figure}[htp]
 \centering
\mbox{ \epsfig{file=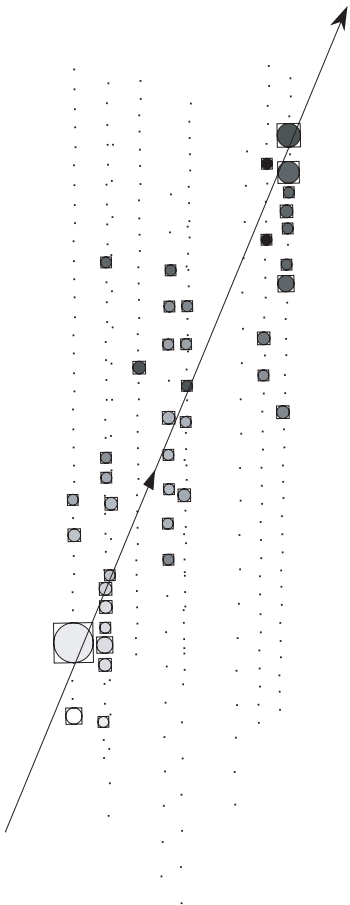,height=12cm}}
\caption{ 
Event display of an upgoing muon event. 
The grey scale 
indicates the flow of time, with early hits at
the bottom and the latest hits at the top of the array.  The arrival times
match the speed of light.  
The sizes of the circles corresponds to the measured amplitudes.
}
\label{fig:event_display}
\end{figure}

We construct a probability distribution function describing the expected
distribution of arrival times, and calculate the likelihood 
$\mathcal{L}_{\mathrm{time}}$
of a given reconstruction hypothesis as the product of the probabilities of
the observed arrival times in each hit OM:
\begin{equation}
\mathcal{L}_{\mathrm{time}} =
 \prod_{i=1}^{N_{\mathrm{hit}}} p(t_{ \mathrm{res}}^{(i)} \, | \,
        d_\perp^{(i)}\!, \theta_{ \mathrm{ori}}^{(i)})
\label{eq:likelihood}
\end{equation}
where $t_{\mathrm{res}} = t_{\mathrm{obs}} - t_{\mathrm{Cher}}$ is the time 
residual (the delay of the observed hit time relative
to that expected for unscattered propagation of Cherenkov
photons emitted by the muon), and $d_\perp$ and $\theta_{\mathrm{ori}}$ are 
the distance of the OM from
the track and the orientation of the module with respect to the track.  The
probability distribution function $p$ includes the 
effects of scattering and absorption in the bulk ice
and in the refrozen ice around the modules.  The functional form of 
$p$ is based on a solution to a transport equation of the
photon flux from a monochromatic 
point source in a scattering medium~\cite{WIEBUSCHRECO,pandel}. 
The free parameters of this function are then fit to the expected 
time profiles that are obtained
by a simulation of the photon propagation from muons 
in the ice~\cite{WIEBUSCHRECO,ptd}.  Varying the
track parameters of the reconstruction hypothesis, we find the maximum of
the likelihood function, corresponding to the best track fit for the event.
The result of the fit is described by five parameters:
three ($x,y,z$) to determine a reference point, 
and two ($\theta,\phi$) for the zenith and azimuth of the track direction. 
Figure \ref{fig:event_display} shows an event display of two upgoing muon
events together with the reconstructed tracks. 

\subsection{Quality Parameters}
\label{subsec:cuts}

The set of apparently 
upgoing tracks provided by the reconstruction
procedure exceeds the expected number of upgoing tracks from atmospheric
neutrino interactions by 
one to three orders of magnitude, depending on the
details of the reconstruction algorithm (see Sections \ref{sec:desy} and
\ref{sec:uw}).  In order to reject the large number of ``fake events''
--- events generated by a downgoing muon or 
cascade, but seemingly having an upgoing structure --- we impose
additional requirements on the reconstructed events to obtain a relatively
pure neutrino sample.  These requirements
consist of cuts on observables derived from the reconstruction
and on topological event parameters.  Below, we describe the most relevant
of the parameters used.

\subsubsection{Reduced Likelihood, $L$}

In analogy to a reduced $\chi^2$, we define a reduced likelihood
\begin{equation}
L = \frac{- \ln{\cal L_{\mathrm{time}}}}{N_{\mathrm{hit}} - 5}
\end{equation}
where $N_{\mathrm{hit}} - 5$, the number of recorded hits in the event less the five track
fit parameters, is the number of degrees of freedom.
A smaller $L$ corresponds to a higher quality of the fit.

\subsubsection{Number of Direct Hits, $N_{\mathrm{dir}}$}

The number of direct hits is defined as the number of hits with time delays
$t_{\mathrm{res}} $ smaller than a certain value.  
We use time intervals of \mbox{[-15\,ns,+25\,ns]}
and \mbox{[-15\,ns,+75\,ns]}, and denote the corresponding parameters as
$N_{\mathrm{dir}}^{(25)}$ and $N_{\mathrm{dir}}^{(75)}$, respectively. 
The negative extent of the window allows for jitter in PMT
rise times and for small errors in geometry and calibration, while the
positive side includes these effects as well as delays due to scattering of
the photons. Events with many
direct hits (i.e., only slightly delayed photons) are likely to be well
reconstructed.  

\subsubsection{Track Length, $L_\mathrm{{dir}}$}

The track length is defined by projecting each of the direct hits onto the
reconstructed track, and measuring the distance between the first and the
last hit.  A cut on this parameter rejects events with a small lever arm
for the reconstruction.  Direct hits with time residuals of
\mbox{[-15\,ns,+75\,ns]} are used for the measurement of the track length.
Cuts on the absolute length, as well as zenith angle dependent cuts (which
take into account the cylindrical shape of the detector) have been used.
The requirement of a minimum track length
corresponds to imposing a muon energy threshold.  For example, a track
length of 100 m translates into a muon energy threshold of about 25 GeV.

\subsubsection{Smoothness, $S$ \label{sec:smoothness}}

\begin{figure}[htp]
\centering
\mbox{ \hskip-1.5cm \epsfig{file=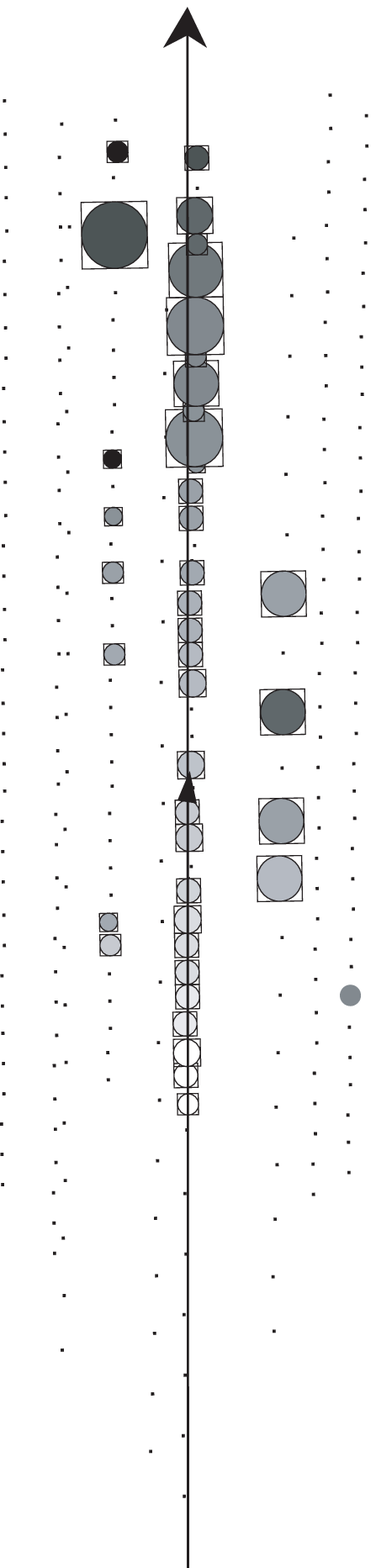,height=12cm}}
\mbox{ \hskip1.0cm  \epsfig{file=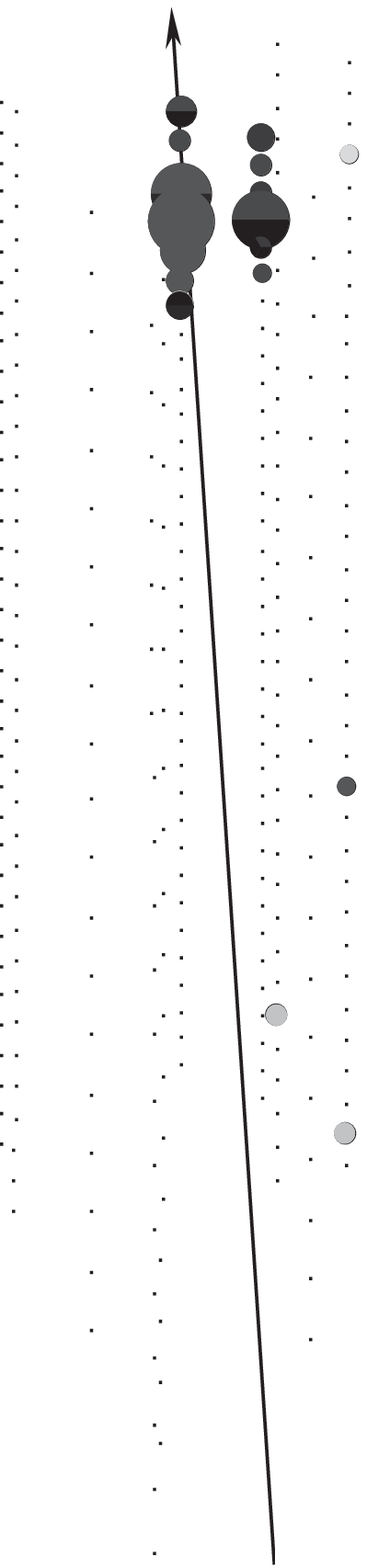,height=12cm}}
\caption{ 
Two muon events: The upgoing muon event shown on the left has a smooth distribution of hits 
along the track.  The track-like hit topology of this event 
can be used to distinguish it from background events. 
The event on the right is a background event with a poor smoothness value. 
}
\label{fig:event_display2}
\end{figure}

The ``smoothness'' parameter is a check on the self-consistency of the
fitted track.
It measures the constancy of light output along the track.  Highly variable
apparent emission of light usually indicates that the track either has been
completely misreconstructed or that an underlaying muonic Cherenkov 
light was obscured by a very bright stochastic light emission, which usually
leads to poor reconstruction.  The
smoothness parameter was inspired by the Kolmogorov-Smirnov test of the
consistency of two distributions; in our case the consistency of the
observed hit pattern with the hypothesis of constant light emission by a
muon.

Figure \ref{fig:event_display2} shows two events to illustrate 
the characteristics of the smoothness parameter. One event is 
a long uniform track, which was well reconstructed. The other event is 
a background event which displays a very poor smoothness. 

\begin{figure}[htb]
\centering
\includegraphics[width=0.95\linewidth]{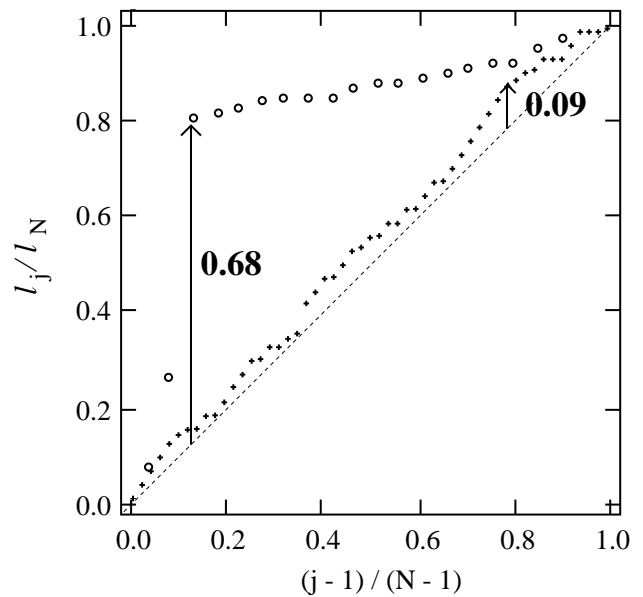}
\caption{Illustration of the smoothness parameter, which compares the
observed distribution of hits to that predicted for a muon emitting
Cherenkov light.  In a simplest formulation, shown here, the prediction is
given as a straight line. A large deviation from a straight line (0.68) 
is found for the event on the right in Fig.~\ref{fig:event_display2}. 
The high quality track-like event on the left in Fig.~\ref{fig:event_display2} 
displays a small deviation (0.09).}
\label{fig:smooth_diagram}
\end{figure}

The simplest definition of the smoothness is given by
\begin{equation}
S = \mathrm{max}(|S_j|) \hspace{5mm} \mbox{where} \hspace{5mm} S_j = \frac{j -1}{N
-1} - \frac{l_j}{l_N}.
\end{equation}
Figure~\ref{fig:smooth_diagram} illustrates the smoothness parameter 
for the two events displayed in Fig.~\ref{fig:event_display2}.
Here $l_j$ is the distance along the track between the points of closest
approach of the track to the first and the $j^{\mathrm{th}}$ hit modules, with 
the hits taken in order 
of their projected position on the track.  $N$ is the
total number of hits.  Tracks with hits clustered at the beginning or end
of the track have $S_j$ approaching $+1$ or $-1$, leading to $S=1$.  High quality
tracks such as the event on the left side of Fig.~\ref{fig:event_display2}, 
with $S$ close to zero, have hits equally spaced along the track.

\subsubsection{Sphericity} 

Treating the hit modules as point masses, we can form a tensor of inertia
for each event, describing the spatial distribution of the hits. 
Diagonalizing the tensor of inertia yields as eigenvalues $I_i$ the moments
of inertia about the principal axes of rotation.  For a long, cylindrical
distribution of hit modules, two moments will be much larger than the
third.  We can reject spherical events, such as those produced by muon
bremsstrahlung, by requiring that the normalized magnitude of the smallest
moment, $I_1/\sum I_i$, be small.

\subsection{Principal Methods of the Analyses}

The two analyses of the data diverge after the filtering stage, following
different approaches to event reconstruction and background rejection.

Analysis I uses an improved likelihood function based on a more detailed
description of the photon response~\cite{WIEBUSCHRECO}, followed by a set
of stepwise tightened cuts.  Analysis II uses a Bayesian
reconstruction~\cite{BAYES} in which the likelihood is multiplied by a
zenith angle dependent prior function, resulting in a strong rejection
of downgoing background.

Rare backgrounds due to unsimulated instrumental effects, such as cross-talk
between signal
channels and unstable voltage supply, were identified in the course of the
analyses.
These effects either produced spurious triggers, or, more often, spurious
hits that caused the event to be misreconstructed.
Different but comparably efficient techniques were developed to treat these
backgrounds.  In Analysis I the event topology is inspected; if the spatial
pattern of hit OMs is inconsistent with the reconstructed muon trajectory,
the event is rejected.  Analysis II attempts to remove the anomalous hits
or triggers through identification of characteristic correlations in signal
amplitudes and times, which 
considerably reduces the rate of these misreconstructions.

At this stage the data set in each analysis is reduced to several
thousand events out of the original $1.05\cdot10^9$, but the data are still
background dominated.  The prediction for atmospheric neutrinos is about
500 at this point.

For the final selection of a nearly pure sample of neutrino induced events,
cuts on characteristic observables are tightened until the remaining
background disappears.  
The two analyses
use different techniques to choose their final cuts, but obtain comparable
efficiencies. Further details of the analyses can be found in
references \cite{BIRON, GAUGDIPL, DeYoung}.

\section{Analysis I \label{sec:desy}}

In this analysis the data were
processed through three levels of initial cuts,
designed to reduce the number of background events to a manageable size for 
the final cut evaluation.
After a first filtering based on the line fit (level~1), cuts on the
zenith angle, the number of direct photons, and
the likelihood of the
fitted track obtained by the maximum likelihood reconstruction were applied
(level~2).

\subsection{Removal of Cascade-Like Events and Detector Artifacts \label{sec:remove_cascade} } 

A third filter level used the results of an iterative likelihood
reconstruction with varying track initializations, a fit based on the hit
probabilities (see Eq.~\ref{eq:phitpnohit}) and a reconstruction to
the hypothesis of a high energy cascade, e.g. due to a 
bright seconday muon bremsstrahlung interaction.

\begin{figure}
\centering
\includegraphics[width=1.0\linewidth]{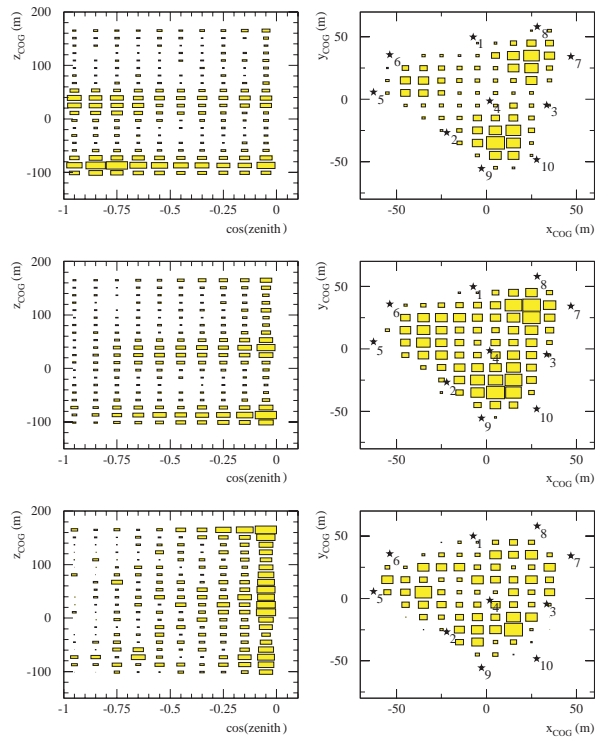}
\caption{Characteristic distributions 
of the center of gravity (COG) of events.  The figures on the
left show the distribution of the depth $z_{\mathrm{COG}}$  versus the 
reconstructed
zenith angle.  The figures on the right show the horizontal location of
events in the $x_{\mathrm{COG}} $-$y_{\mathrm{COG}} $ plane of events 
with $0$ m $ < z_{\mathrm{COG}} <
50$ m.  The positions of the strings are marked by stars.  Top: Experimental
data before application of the COG cuts.  Middle: Experimental data after
application of the COG cuts.  Bottom: Expectation from the BG simulation
after cuts.
\label{abb:whycog}
}
\end{figure}

The first two levels of filtering consisted of relatively weak cuts on
basic parameters like the zenith angle and likelihood.  They reduced the
data set to about $4\cdot 10^{5}$ events.
At this stage, residual unsimulated instrumental
features become apparent, e.g., comparatively high amplitude cross-talk 
produced when 
a downgoing muon emits a bright shower in the center
of the detector.  Such events are predominantly reconstructed as moving
vertically upward and can be
identified in the distribution of the center of gravity (COG) of hits. 
Its vertical component ($z_{\mathrm{COG}}$) shows unpredicted peaks in the 
middle and the
bottom of the detector
(see also Fig.~\ref{fig:cogzfinal} (top), demonstrating the effect for 
Analysis II), 
while the horizontal components ($x_{\mathrm{COG}}$ and
$y_{\mathrm{COG}}$) show an enhancement of hits towards the outer strings.  
These strings
are read out via twisted pair cables, as opposed to the coaxial
cables used on the inner strings.  The twisted pair cables were found to
be more susceptible to cross-talk signals. Note that
variations in the optical parameters of the ice due to past climatological
episodes also produce some vertical structure.  

We developed additional COG cuts on the topology of the events in
order to remove these backgrounds.  
These cuts, which depend on the reconstructed zenith angle, use the track
lengths $L_{\mathrm{dir}}$ and the normalized smallest eigenvalues of the tensor of 
inertia ($I_1/\sum I_i$).

Figure~\ref{abb:whycog} shows the different components of the center of
gravity of the hits and the reconstructed zenith angle before and after
application of the COG cuts, and the Monte Carlo prediction for
fake upward events stemming from misreconstructed 
downgoing muons.  The cuts remove most of the unsimulated background
-- in particular that far from the horizon -- 
and bring experiment and simulation into much better agreement.

In order to verify the signal passing rates, these cuts and those from the
previous levels were applied to a subsample of unfiltered (i.e. downgoing)
events but with the zenith angle dependence of the cuts reversed, thus
using the abundant cosmic ray muons as stand-ins for upgoing muons. 

In all, 
these three levels of filtering reduced the data set by
a factor of approximately 
$10^{5}$ (see Table~\ref{tab:passrates}).

\subsection{Multi-Photoelectron Likelihood and Hit Likelihood}

Before the final cut optimization the last, most elaborate 
reconstruction was applied,
combining the likelihoods for the arrival time of the first 
of muliple photons in a PMT with the likelihoods for PMTs
to have been hit or have not been hit.

The probability densities
$p(t_{ \mathrm{res}}^{(i)} \, | \,
        d_\perp^{(i)}\!, \theta_{ \mathrm{ori}}^{(i)})$ (see 
eq.1, Section \ref{sec:max_likelihood}) describe only the arrival times of single
photons. Density functions for the multi-photoelectron case
have to include the effect of repeatedly sampling 
the distribution of photon arrival times.
For several detected photons, the first of them is usually
less scattered
than the average photon (which defines the
single photoelectron case). Therefore the leading edge of a PMT
pulse composed of multiple photoelectrons (MPE) will be 
systematically shifted to
earlier times compared to a single photoelectron.  
The {\it MPE likelihood} 
${\cal{L}}_{\mathrm{time}}^{\mathrm{MPE}}$~\cite{WIEBUSCHRECO} 
uses the recorded amplitude information to model this shift.

In the reconstructions mentioned so far, the timing information
from hit PMTs was used. However, a PMT which was {\it not}
hit also delivers information. The {\it hit likelihood}
$\cal{L}_{\mathrm{hit}}$ does not depend on the
arrival times but represents the probability that the track 
produced the observed hit pattern.
It is constructed from the probability densities
$p_{\mathrm{hit}}(d_\perp^{(i)}\!, \theta_{\mathrm{ori}}^{(i)})$
that a given PMT $i$ was hit if it was in fact hit, and the probabilities
$(1 - p_{\mathrm{hit}}(d_\perp^{(j)}\!, \theta_{\mathrm{ori}}^{(j)})$
that a given PMT $j$ was not hit if it was not hit:

\begin{eqnarray}
\begin{array}{l}
{\cal L}_{\mathrm{hit}} = \\
\prod_{i=1}^{N_{\mathrm{hit}}}
p_{\mathrm{hit}}(d_\perp^{(i)}\!, \theta_{\mathrm{ori}}^{(i)}) \;
 \cdot 
\prod_{i=N_{\mathrm{hit}}+1}^{N_{\mathrm{OM}}} \left( 1 - 
 p_{\mathrm{hit}}(d_\perp^{(i)}\!, \theta_{\mathrm{ori}}^{(i)}) \;  \right)
\label{eq:phitpnohit}
\end{array}
\end{eqnarray}

\noindent
where the first product runs over all hit PMTs
and the second over all non-hit PMTs.

The likelihood combining these two probabilities is 

\begin{equation}
 {\cal L} = {\cal L}_{\mathrm{time}}^{\mathrm{MPE}} 
\cdot {\cal L}_{\mathrm{hit}} ~
\label{eq:mpe}
\end{equation}

A cut on the
reconstructed zenith angle obtained from fitting with
${\cal L}$ leaves less than $10^{4}$ events in
the data set, defined as level 4 in Fig. \ref{abb:passrates}.

\subsection{Final Separation of the Neutrino Sample
 \label{sec:cuteval}}

\newcommand{\taboptions}{htbp}
\begin{table*}[\taboptions]
\newcommand{\m}{\hphantom{$-$}}
\newcommand{\cc}[1]{\multicolumn{1}{c}{#1}}
\renewcommand{\tabcolsep}{1pc} 
\renewcommand{\arraystretch}{1.1} 
\begin{tabular}{|l|l||l|}
\hline
Parameter & Cut & Explanation \\ \hline \hline
$|S|$ & $< 0.28 $ & see Section~\ref{sec:smoothness} \\ \hline
$|S^{P_{\mathrm{hit}}}|/(\theta_{\mathrm{mpe}} - 90^\circ)$  & $< 0.01$ & 
                          Tightens the requirement on the  smoothness for tracks \\
         & &              close to the horizon where background is high.  \\ \hline
$(N_{\mathrm{dir}} - 2) \cdot  L_{\mathrm{dir}}$ & $> 750$m & 
                            Lever arm of the track times the number of supporting points. \\ 
                                       \hline
$\log{({\cal L}_{\mathrm{up}}/{\cal L}_{\mathrm{down}})}$ & $< -7.7$ & Ratio of
the likelihoods of the best \\ & & upgoing and best downgoing hypotheses. \\
\hline 
$\Psi(\mbox{mpe,lf})$ & $< 35^\circ$ & Space angle
between the results from the multi-photon                     \\ 
 &  & likelihood reconstruction and the line fit.  This cut   \\ 
 &  & effectively removes cross-talk features.                \\ \hline
$\sqrt{(S_{\mathrm{dir}})^2+(S^{P_{\mathrm{hit}}}_{\mathrm{dir}})^2} $ & $< 0.55$ & Parameter
combining the two smoothness definitions (here                \\
 &  & calculated using only direct hits). This cut effectively removes  \\ 
 &  & coincident muon events from independent air showers.  \\ \hline
\end{tabular}
\caption{Final quality parameters and cuts obtained from
the cut evaluation procedure.  The ``direct'' time interval for variables
$N_{\mathrm{dir}}$, $L_{\mathrm{dir}}$, and $S_{\mathrm{dir}}$ is \mbox{[-15\,ns,+75\,ns]}.  The first
four rows show cut parameters obtained by all
(Monte Carlo and experimental) searches,
the last two rows show two additional (weaker) cuts, which were found to
remove unsimulated backgrounds.  }
\label{tab:cutsI}
\end{table*}

For the final stage of filtering, a method (\cuteval) was developed to
select and optimize the cuts taking into account correlations between the
cut parameters.  A detailed description of this method can be found
in~\cite{GAUGDIPL}.  The principle of \cuteval is to numerically optimize
the ratio of signal to $\sqrt{\text{background}}$ by variation of the
selection of cut parameters, as well as the actual cut values.  Parameters
are used only if they improve the efficiency of separation over optimized
cuts on all other already included parameters.  A first optimization was based purely on Monte
Carlo, with simulated atmospheric neutrinos for signal and simulated
downgoing muons forming the background.  This optimization yielded four
such independent parameters.  Two other optimizations involved experimental
data.  In both cases, experimental data 
have been defined as the
background sample.  In one case, the signal was represented by
atmospheric neutrino Monte Carlo, in the other by experimental data
subjected to zenith angle inverted cuts 
(i.e. to downward events passing
the quality cuts, but being ``good'' events with respect to the upper
hemisphere instead -- like neutrino candidates -- with respect to the lower
hemisphere). 
 These latter optimizations yielded two additional parameters, 
which  rejected a small
contribution of residual unsimulated backgrounds: coincident muons from
simultaneous independent air showers and events accompanied by instrumental
artifacts such as cross-talk.  After application of these two cuts to
simulated and experimental data, the distributions of observables agree to
a satisfactory precision.

\begin{figure}
\centering
\mbox{\epsfig{file=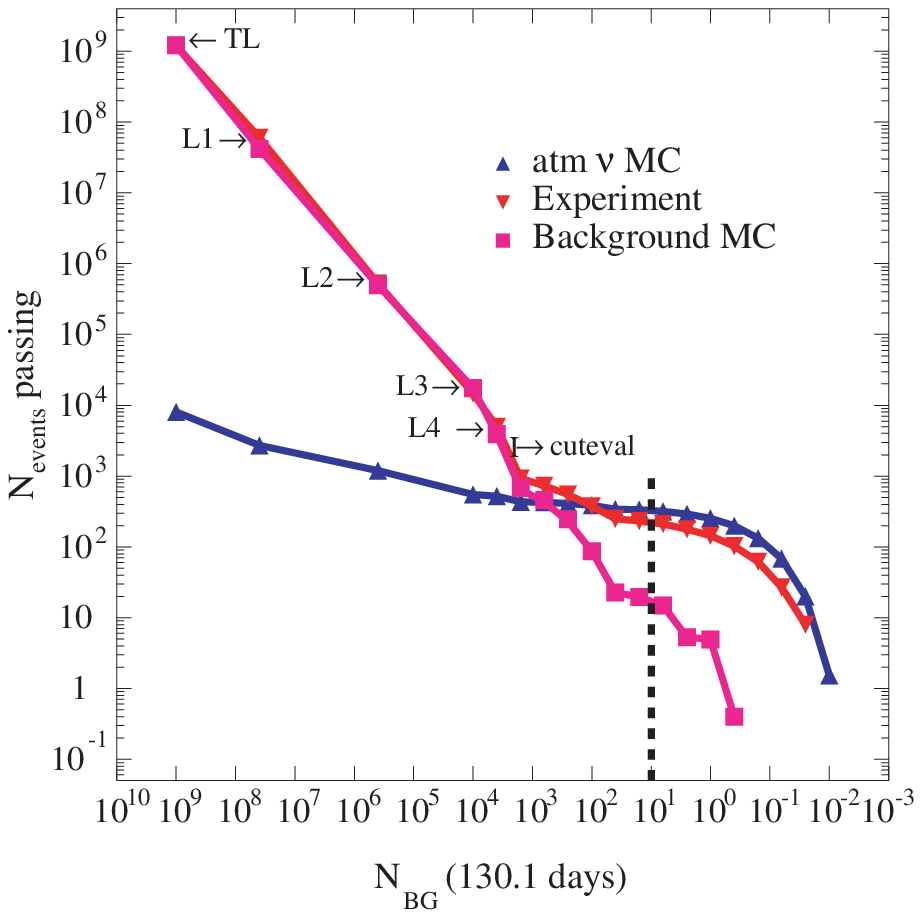,width=7.6cm}}
\mbox{\epsfig{file=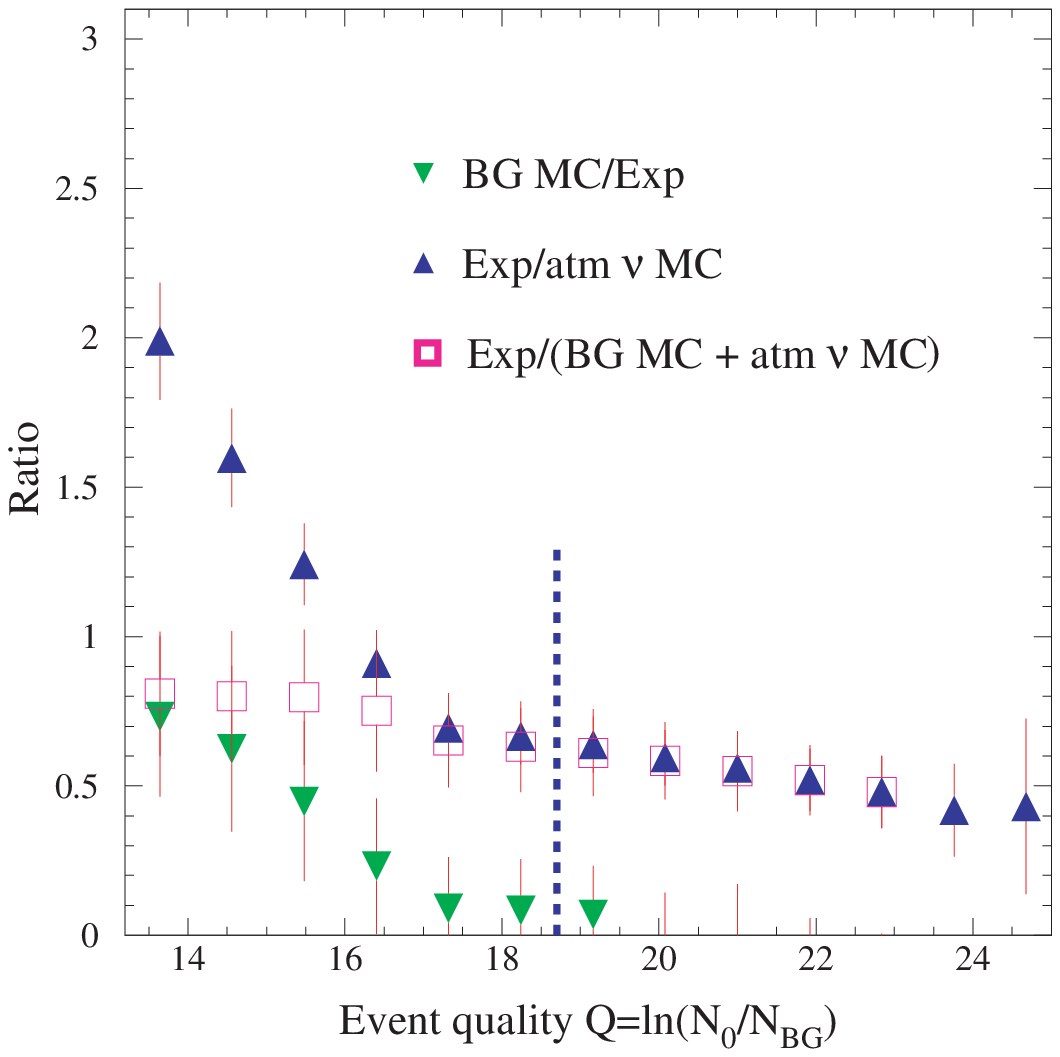,width=7.6cm}}
\caption{
 The fitted background parameter $N_{\mathrm{BG}} $.  Top: Number of events versus
 $N_{\mathrm{BG}} $.  Smaller values of $N_{\mathrm{BG}} $ correspond to harder
 cuts.  Below $N_{\mathrm{BG}} = 1500 $ the \cuteval parameterization was used to
 calculate the cut values corresponding to $N_{\mathrm{BG}} $.  For larger values of
 $N_{\mathrm{BG}} $ the data points correspond to the cuts from the filter levels:
Level 4 (see Section V B), Level 3, Level 2, Level 1, and trigger level
 (table~\ref{tab:passrates}). Bottom: Ratios of
events passing in the experimental data
 compared to various Monte Carlo
 expectations for signal and background as a function of event quality.
 The dashed line indicates the final cuts.}
\label{abb:passrates}
\end{figure}

Once the minimal set of parameters is found, the optimal cut values
can be represented as a function of the number of
background events $N_{\mathrm{BG}}$ passing the cuts.
The result is a path through the cut parameter space 
which yields the best signal efficiency for 
any desired purity of the signal, characterized by $N_{\mathrm{BG}}$.
Using this representation, one can
calculate the number of
events passing the cuts as a function of the fitted $N_{\mathrm{BG}} $ for
signal and for background Monte Carlo.
Figure~\ref{abb:passrates} (top) shows this dependence
for simulations as well as for experimental data,
with $N_{\mathrm{BG}} $ varying from trigger level to a
level that leaves only a few events in the data set.
One observes that the actual background expectation falls roughly linearly
as the fitted $N_{\mathrm{BG}} $ is reduced.
Below values of a few hundred events the signal is expected to dominate the
event sample.  The experimental curve follows the expectation from the sum of
background and signal
Monte Carlo.
For large $N_{\mathrm{BG}} $, the observed event
rate follows the background expectation.  At smaller
$N_{\mathrm{BG}} $, the
experimental shape turns over into the signal expectation and follows it
nicely down to the sample of events with highest quality
(the smallest values
of $N_{\mathrm{BG}} $).
For a moderate background contamination of
$N_{\mathrm{BG}} = 10$, one gets a total of 223 neutrino
candidates.  The parameters and cut values as obtained by the
\cuteval procedure are summarized in Table~\ref{tab:cutsI}.

Figure~\ref{abb:passrates} (bottom) translates the background parameter
$N_{\mathrm{BG}}$ into
an event quality parameter $Q$, defined as $Q~\equiv~ \ln ( { N_{0}/N_{\mathrm{BG}} }
)~=~ \ln ( { 1.05\cdot 10^{9}/N_{\mathrm{BG}} } )$.  The plot shows the ratios of
events from the upper figure as a function of $Q$.  At higher qualities ($Q
> 17$), the ratio of observed events to the atmospheric neutrino simulation
flattens out
with a further variation of only 30\%.  The value at $Q=17$ is
approximately unity for the {\tt angsens} Monte Carlo and
about 0.6 for the standard Monte Carlo
(chosen in Fig.~\ref{abb:passrates}, top) and approximately unity
  for the angsens Monte Carlo (chosen in Fig.~\ref{abb:passrates}, bottom).


\begin{table*}[\taboptions]
\newcommand{\m}{\hphantom{$-$}}
\newcommand{\cc}[1]{\multicolumn{1}{c}{#1}}
\renewcommand{\tabcolsep}{1pc} 
\renewcommand{\arraystretch}{1.0} 
\begin{tabular}{|l|c|c|c|c|}
\hline
Filter level & Atm.  $\nu$ & Atm.  $\nu$ MC  & Atm.  $\mu$ MC &
Experimental\\
 &     MC     & {\tt angsens}  & (Background) &   data\\
\hline
Events at trigger level   & 8978 & 5759 & $9.03\cdot 10^{8}$ & $1.05\cdot
10^{9}$\\
\hline
Efficiency at Level 1 & 0.34 & 0.37 & 0.4 $\cdot 10^{-1}$ & 0.5 $\cdot10^{-1}$ \\ 
Efficiency at Level 2 & 0.15 & 0.15 & 0.4 $\cdot 10^{-3}$ & 0.4 $\cdot 10^{-3}$ \\ 
Efficiency at Level 3 & 0.7 $\cdot 10^{-1}$ & 0.7 $\cdot
10^{-1}$ & 0.7 $\cdot 10^{-5}$ & 0.1 $\cdot 10^{-4}$ \\
Efficiency after final cuts & 0.4 $\cdot 10^{-1}$ & 0.4 $\cdot 10^{-1}$ & 0.6 $\cdot
10^{-8}$ & 0.2 $\cdot 10^{-6}$ \\
\hline
No. of events  & $362 \pm 4$ & $237 \pm 6$ & $7\pm 5$ & 223 \\ 
passing final cuts & & & normalized & \\
\hline
\end{tabular}
\caption{
The cut efficiencies for
the atmospheric neutrino Monte Carlo (MC), the atmospheric muon background Monte
Carlo, and the experimental data for 130 days of detector livetime.
Efficiencies are given for filter levels L1 to L4. L4 is the final selection. 
All errors are purely statistical. The final background
prediction of 7 events has been normalized at trigger level.}
\label{tab:passrates}
\end{table*}

Table~\ref{tab:passrates} lists the cut efficiencies for the atmospheric
neutrino simulation (with and without the implementation of the angular
sensitivity fitted model {\tt angsens}  of the OMs --- see
Sections~\ref{sec:data} and~\ref{sec:syst}), the background
simulation of atmospheric muons from air showers  
(without {\tt angsens}) 
and the experimental data. 
Again, the experimental numbers agree well with the
background simulation up to the first two filter levels.  Later, the Monte
Carlo underestimates the experimental passing rates slightly.  The last row
shows the expected numbers of events for the last stage of filtering. 
If, in addition, the effect of neutrino oscillations (see
Section~\ref{sec:syst}) is included, the atmospheric neutrino simulation
including the {\tt angsens} model predicts
224
events, in closest agreement with the experiment. 
 However, the 5\% effect
due to oscillations is smaller than our systematic uncertainty (see
Section~\ref{sec:syst}).

\subsection{Characteristics of the Neutrino Candidates}

\subsubsection{Time Distribution}

\begin{figure}
\centering
\includegraphics[width=.95\linewidth]{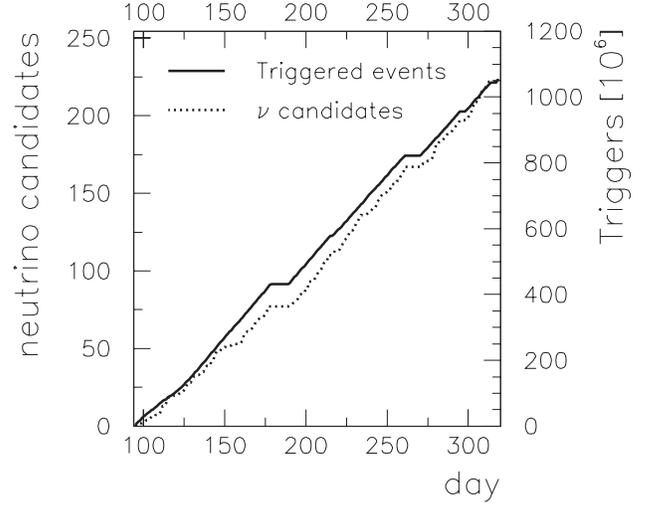}
\caption{The integrated exposure of the AMANDA detector in 1997.
The figure shows the cumulative number of triggers (upper curve) and the
number of observed neutrino events (lower curve) versus the day number.
The intervals with zero gradient correspond to periods where the detector was
not operating stably; data from these periods were excluded from the
analysis.}
\label{fig:lumiplot}
\end{figure}

Figure~\ref{fig:lumiplot} shows the
cumulative number of neutrino events as well as the cumulative
number of event triggers plotted versus the day number in 1997.
One can observe that the neutrino events follow
the number of triggers, albeit with a small deficit during the
Antarctic winter.
This deficit is consistent with statistical fluctuations.
(Actually, seasonal variations slightly {\it de}crease the
downward muon rate during the Antarctic winter \cite{Adam}
and should result in a 10\% deficit of triggers
with respect to upward neutrino events.)

\subsubsection{Zenith Angle Distribution}

\begin{figure}
\centering
  \mbox{\epsfig{file=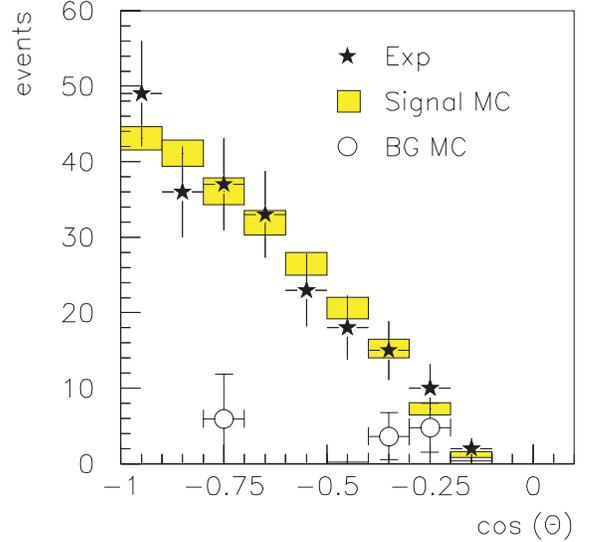,width=7.5cm}}
\caption{Zenith angle distribution
of the experimental data
compared to simulated atmospheric neutrinos
and a simulated background of downgoing muons produced by
cosmic rays.  In this figure the Monte Carlo prediction is
normalized to the experimental
data. The error bars report only statistical errors.
}
\label{fig:zenith_ana1}
\end{figure}

Figure~\ref{fig:zenith_ana1} shows the zenith angle distribution of the 223
neutrino candidates
 compared to the Monte Carlo prediction for atmospheric neutrinos
\cite{nusim} and the few remaining events predicted by background
simulations. Note that Monte Carlo is normalized to experiment. (The total
number of events is 362 for the atmospheric neutrino simulation
and 223 for experiment, i.e. there is a deficit of 39 percent in
the absolute number of events.)
There is good agreement between
the prediction and the experiment in the shape of the
angular distribution.

\subsubsection{Characteristic Distributions and Visual Inspection \label{sec:nu_character} }

\begin{figure}[htbp]
\centering
   \mbox{\epsfig{file=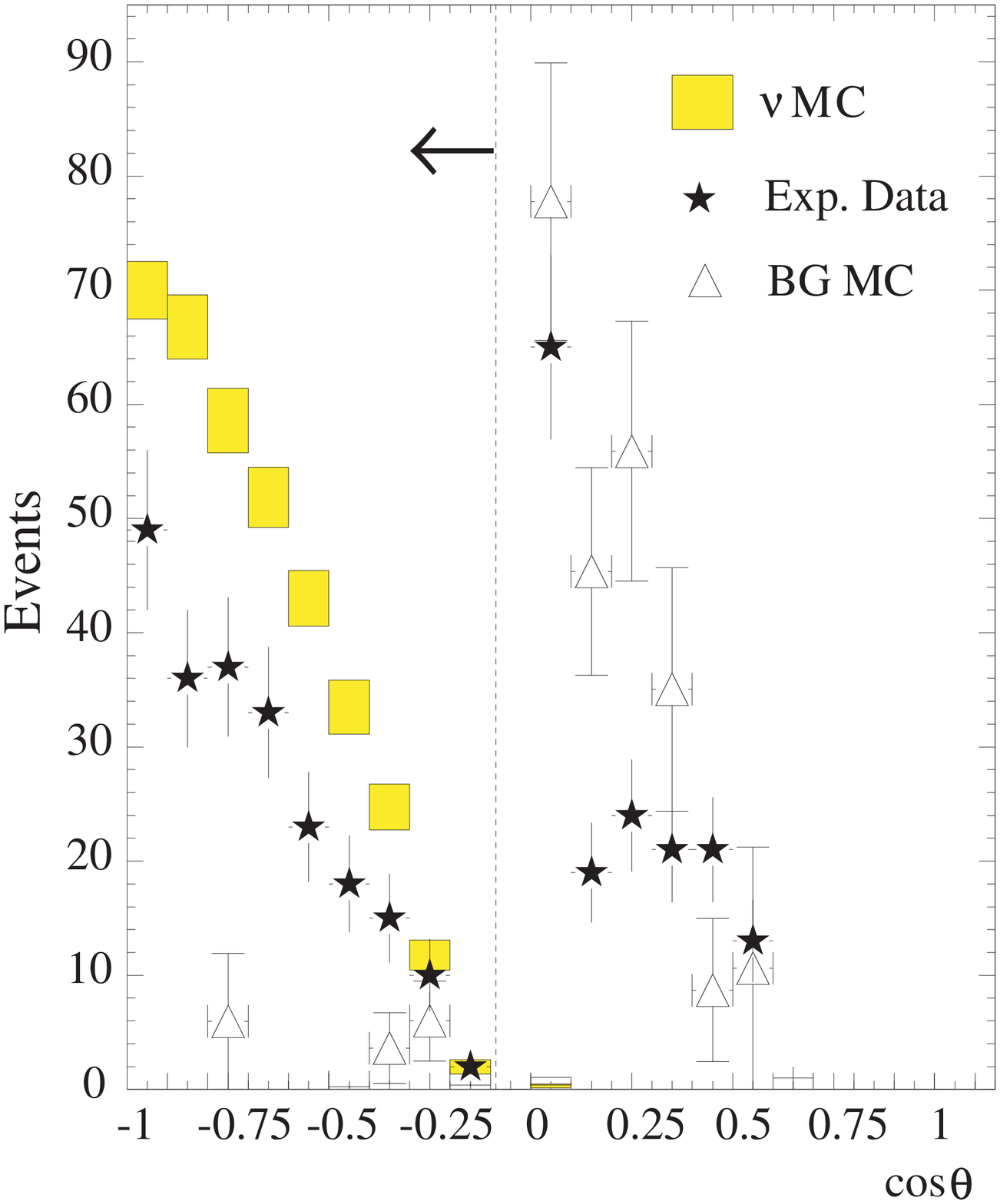,width=0.8\linewidth}}
   \mbox{\epsfig{file=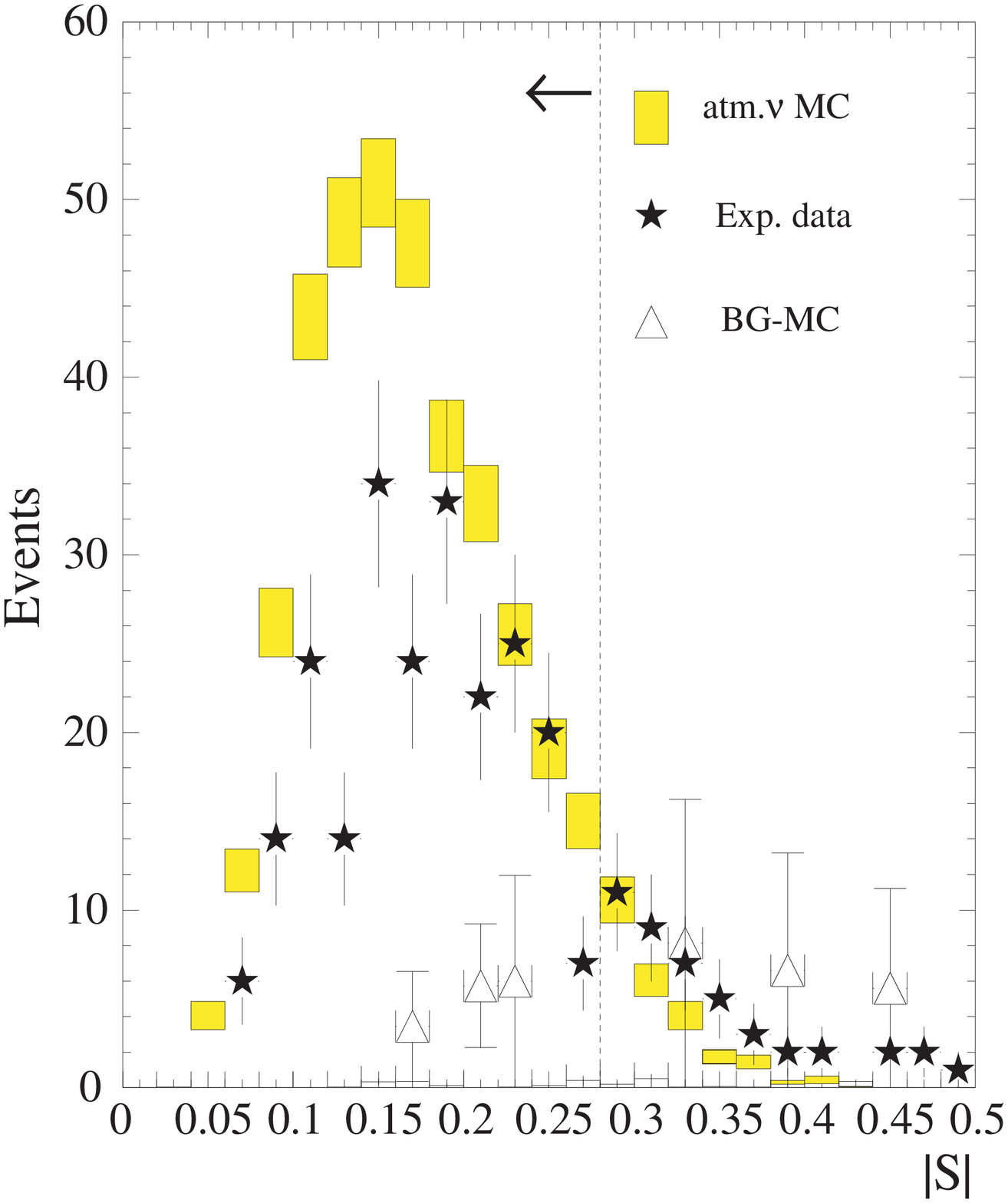,width=0.8\linewidth}}
\caption{ Two
distributions of variables used as cut parameters in the last filter level
(see Table 1 for an explanation of the
variables).  In both cases, all final cuts with the exception of the
variable plotted have been applied.  The
cuts on the displayed parameters are indicated
by the dashed vertical lines. Arrows indicate the accepted parameter space.}
\label{fig:n-1_cuts}
\end{figure}

Four methods
were used to evaluate the effectiveness of
the analysis and the level of residual backgrounds:
a)~\emph{$N\!-\!1$~cuts}, b)~\emph{unbiased~variables},
c)~\emph{low~level~distributions}, and d)~\emph{visual~inspection}.\\ \\
a) The \emph{$N\!-\!1$ test} evaluates the $N$ final cuts
one by one and yields an estimate
of the background contamination in the final sample.  One applies all but
one of the final cuts (the one in the selected variable),
and plots the data in this variable.
In the signal region of this variable (defined by the later
applied cut) shapes of experiment and signal Monte Carlo should agree.
In the background region, the experimental data should approach
the expected background shape.
Figure~\ref{fig:n-1_cuts} shows four of these distributions.
The applied cut is shown by a dotted line.
All four cuts satisfy the
test: the shape of the distributions agree reasonably well
on both sides of the applied cuts. 
Two $N-1$ distribution from Analysis II are shown in Fig.~\ref{fig:nminus1plots}. \\ \\
b) An obvious test is the investigation of distributions of \emph{unbiased
variables} (i.e. variables to which no cuts have been
applied) in the final neutrino sample.  Here, the experimental 
distributions follow the Monte Carlo
signal expectations nicely.  Some deviations are observed, especially
in the number of OMs hit
and the 
velocity $v_{\mathrm{lf}}$ obtained from the line fit (see Section IV,A).
However, as can bee seen from Fig.~\ref{fig:linefit},
part of these disagreements disappear if the
the standard atmospheric neutrino MC is replaced by
the {\tt angsens} MC version. \\ \\
c) In order to account for possible
pathological \emph{low level} features in the
data sample (especially cross-talk), we
{\it i)} investigated basic pulse amplitude and 
pulse width (TOT) distributions and {\it ii)}
re-fitted all events after the cross-talk hit cleaning
procedure applied in Analysis~II (which is tighter than the
standard cross-talk cleaning introduced in Section \ref{sec:clean_filter} ).
Both: these distributions and that for the recalculated
zenith angles show no significant deviation from the previous
ones.  No cross-talk features are found in the resulting neutrino sample.\\ \\
d) Finally, a~\emph{visual inspection} of the full neutrino sample was
performed, by visually displaying each event like in Fig.~4.  
The visual inspection gives consistent results with the other
methods of background estimation and yields an upper limit on the
background contamination of muons from random coincident air showers
(see below).

\subsection{Background estimation \label{sec:bgest}}

The results of four independent methods of background estimation are
summarized in table~\ref{tab:bgest}.

\begin{figure}
\centering
\includegraphics[width=.95\linewidth]{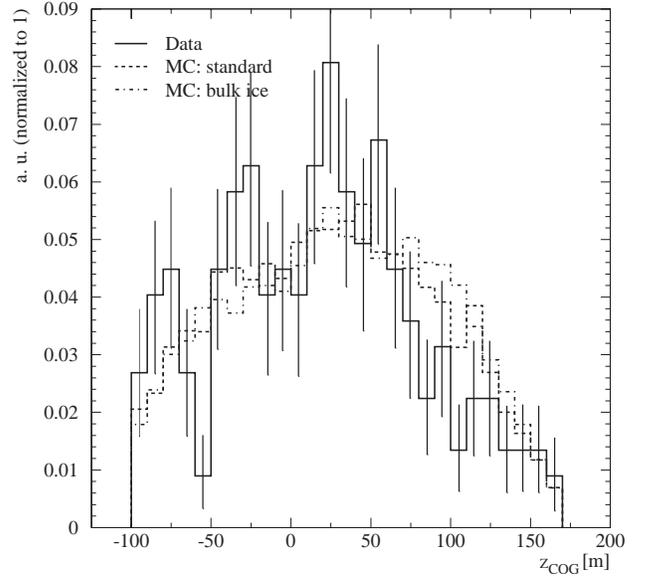}
\caption{Distributions of $z_{\mathrm{COG}} $ for the experiment and
atmospheric neutrino signal Monte Carlo.
{\it MC standard} and {\it MC bulk ice} denote two different
ice models. The first includes vertical ice layers in accordance
with Fig.\,2, the second uses homogeneous ice.
}
\label{abb:cogz}
\end{figure}

First, the background Monte Carlo itself
gives an estimate. It yields 7 events if rates are normalized to
the trigger level (see table~\ref{tab:passrates}).  Because the passing rates
differ slightly between the experiment (higher) and the background Monte Carlo
(lower), we made the conservative choice to renormalize the background Monte
Carlo to the level~3 experimental passing rate.  This gives an estimate of
about 16 background events in our final sample.

\begin{table}[hbtp]
{\small
\begin{center}
\begin{tabular}{|l|c|}
\hline
BG estimation method & estimation \\
\hline
BG MC & $16 \pm 8$  \\
$N\!-\!1$ cuts & $14 \pm 4$ \\
$z_{\mathrm{COG}}$ distributions & $<$ 35\\
Visual inspection &
$<$ 23\\
\hline
\end{tabular}
\end{center}
\caption{\label{tab:bgest} Various
estimates of the background remaining in the
experimental data sample of 223 neutrino candidates.
}}
\end{table}

From the $N\!-\!1$ distributions we obtained an alternative
approximation of the residual background.
We re-normalized both signal and
background MCs in the background region to fit the number of experimental
events in the background region.  The number of re-normalized background MC
events in the signal region is then a background estimate.
This estimate
was performed $N$ times (once for each $N\!-\!1$ distribution).
The average over
all $N$ estimations yields 14 background events.
Note that this
averaging procedure is reasonable only for the case of independent cuts.
With the method by which we have chosen the cut parameters,
this condition is fulfilled to first approximation.

We have found that
cross-talk hits are related to the characteristic triple-peak structure
in the distribution of the vertical component of the center of gravity of
hits ($z_{\mathit{\mathrm{COG}}}$)  
which has been discussed in Section \ref{sec:remove_cascade} -- see Fig.7 and 
also Fig.\ref{fig:cogzfinal} (top).
Since there are remaining
cross talk hits which have survived the standard cleaning
(see section \ref{sec:clean_filter}), this distribution was studied
in detail.
As shown in Fig.~\ref{abb:cogz},
the final experimental sample of neutrino candidates shows no
statistically significant excess with respect to the atmospheric neutrino
Monte Carlo prediction in the regions of the characteristic peaks.
Therefore, an upper limit on this special class
of background was derived and yields $< 35$ events.

The visual inspection of the neutrino sample yields 13 events.
Seven of them show the signature of coincident muons
from independent air showers; i.e., two well separated spatial
concentrations of hits,
each with a downward time flow but with the lower group
appearing earlier than the upper one.
Taking into account the scanning efficiencies which were determined
by scanning signal and background Monte Carlo events,
an upper limit of 23 events is obtained from visual inspection.

Combining the results from the above methods, the expected
  background is estimated to amount to 4 to 10\%
of the 223
experimental events.

\section{Analysis II \label{sec:uw}}

The second analysis follows a different approach;
 instead of
optimizing cuts to reject misreconstructed cosmic ray muons, this analysis
concentrates on improving the reconstruction algorithm
with respect to background rejection. 
The large downgoing muon flux implies that even a small fraction of downgoing muons
misreconstructed as upgoing will produce a very large background rate. 
Equivalently, for each apparently upgoing event, there were many more
downgoing muons passing the detector than there were upgoing muons; even
though any single downgoing muon had only a small probability of faking an
upgoing event, the total probability that the event was a fake is quite
high.

\subsection{Bayesian Reconstruction}

This analysis of the problem motivates a Bayesian approach \cite{BAYES} to
event reconstruction.  Bayes' Theorem in probability theory states that for
two assertions $A$ and $B$,
\[ P(A \, | \, B) \; P(B) = P(B \, | \, A) \; P(A), \]
where $P(A\,|\,B)$ is the probability of assertion $A$ given that $B$ is true. 
Identifying $A$ with a particular muon track hypothesis $\mu$ and $B$ with
the data recorded for an event in the detector, we have
\[ P(\mu \, | \, \text{data} ) = 
\mathcal{L}_{\mathrm{time}}( \text{data}\, | 
\, \mu) \;P(\mu),\]
where we have dropped a normalization factor $P(\text{data})$ which is a
constant for the observed event.  The function 
$\mathcal{L}_{\mathrm{time}}$ is the regular
likelihood function of Eq.~\ref{eq:likelihood}, and $P(\mu)$ is the
so-called prior function, the probability of a muon $\mu = \mu(x, y, z,
\theta, \phi)$ passing through the detector.

For this analysis, we have used a simple one-dimensional prior function,
containing the zenith angle information at trigger level 
in Fig.~\ref{fig:trig_zenith}.  By
accounting in the reconstruction for the fact that the flux of downgoing
muons from cosmic rays is many orders of magnitude larger than that of
upgoing neutrino-induced muons, the number of downgoing muons that are
misreconstructed as upgoing is greatly reduced.  
It should be noted that
the objections that are often raised with respect to the use of Bayesian
statistics in physics are not relevant to this problem: the prior function
is well defined and normalized and independently known to relatively good
precision, consisting only of the fluxes of cosmic ray muons and
atmospheric muon neutrinos.

\subsection{Removal of Instrumental Artifacts}

The Bayesian reconstruction algorithm is highly efficient at rejecting
downgoing muon events.  Of $2.6\cdot10^8$ events passing the fast filter,
only $5.8\cdot10^4$ are reconstructed as upgoing.  By contrast, the
standard maximum likelihood reconstruction produces about $2.4 \cdot 10^7$
false upgoing reconstructions.  However, less than a thousand neutrino
events are predicted by Monte Carlo, so it is clear that a significant
number of misreconstructions remain.

Detailed inspection of the $5.8\cdot10^4$  events reveals that 
the vast majority is
produced by cross-talk overlaid on triggers from downgoing muons emitting
bright stochastic light near the detector.  This cross-talk confuses the
reconstruction algorithm, producing apparently upgoing tracks.  Because
cross-talk is not included in the detector simulation, the characteristics
of the fakes are not predicted well by the simulation, and the rate of
misreconstruction is much higher than predicted.

\begin{figure}
  \centering \includegraphics[width=0.86\linewidth]{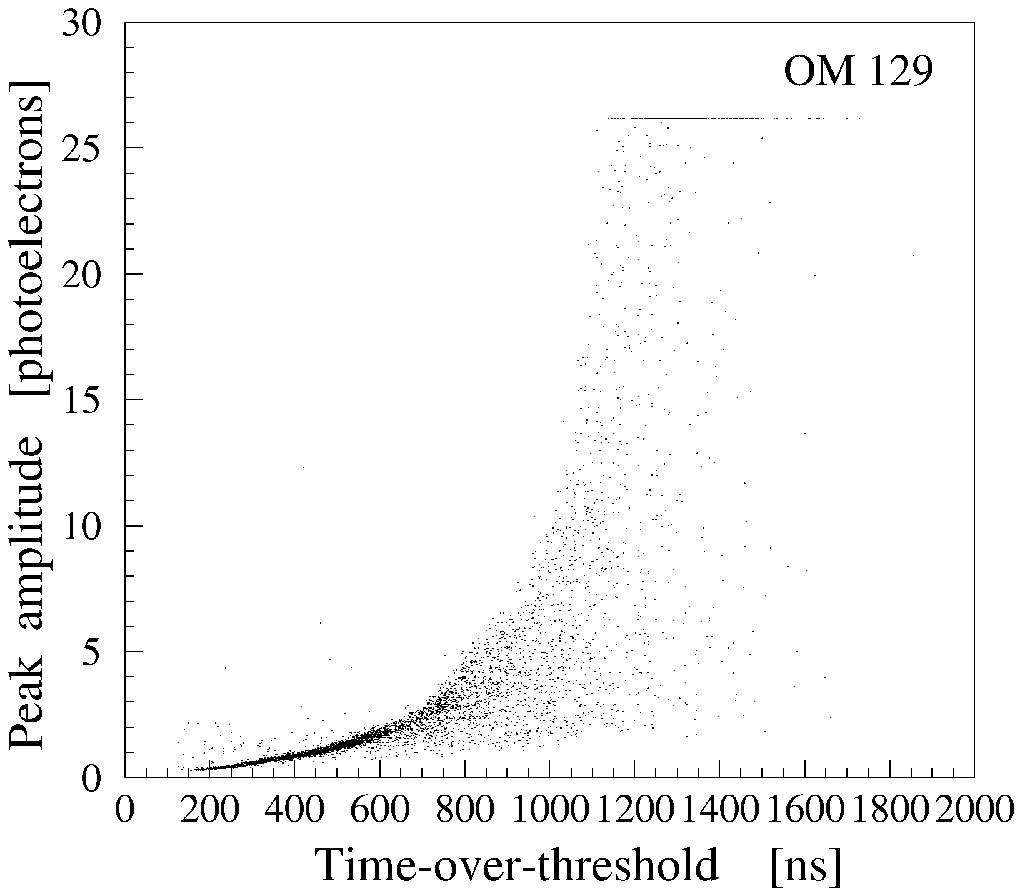}
  \hfill \includegraphics[width=0.86\linewidth]{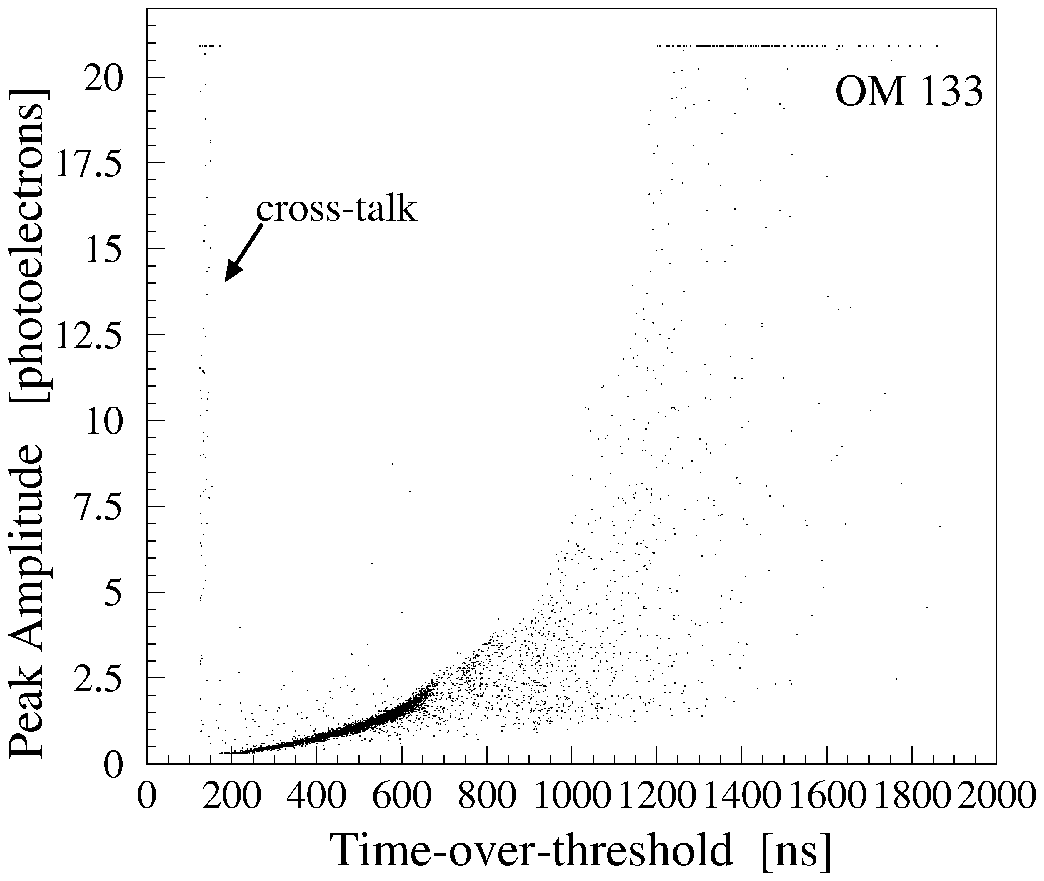}
\caption{Pulse amplitude \emph{vs}.\ duration for modules on the
outer strings.  Normal hits lie in the distribution shown in the upper figure. 
High amplitude pulses of more than a
few photoelectrons are valid only if the pulse width is also large.
Cross-talk induced pulses of high amplitude are characterized by small
time-over-threshold (TOT).  The cut-off seen at high amplitude 
is due to saturation of the amplitude readout electronics.}
\label{fig:adctot}
\end{figure}

The cross-talk is removed by additional hit cleaning routines developed by
examination of this cross-talk enriched data set.  For example, cross-talk
in many channels can be identified in scatter plots of pulse 
width \emph{vs}.~amplitude, 
as shown in Fig.~\ref{fig:adctot}.  The pulse width is measured
as time-over-threshold (TOT).  Real hits form the distribution
shown on the left.  High amplitude pulses should have large pulse width. 
This is not the case for cross-talk induced pulses.  In channels with high
levels of cross-talk, an additional vertical band is found at high
amplitudes but short pulse widths, as seen in the lower figure.

\begin{figure}
  \centering \includegraphics[width=0.85\linewidth]{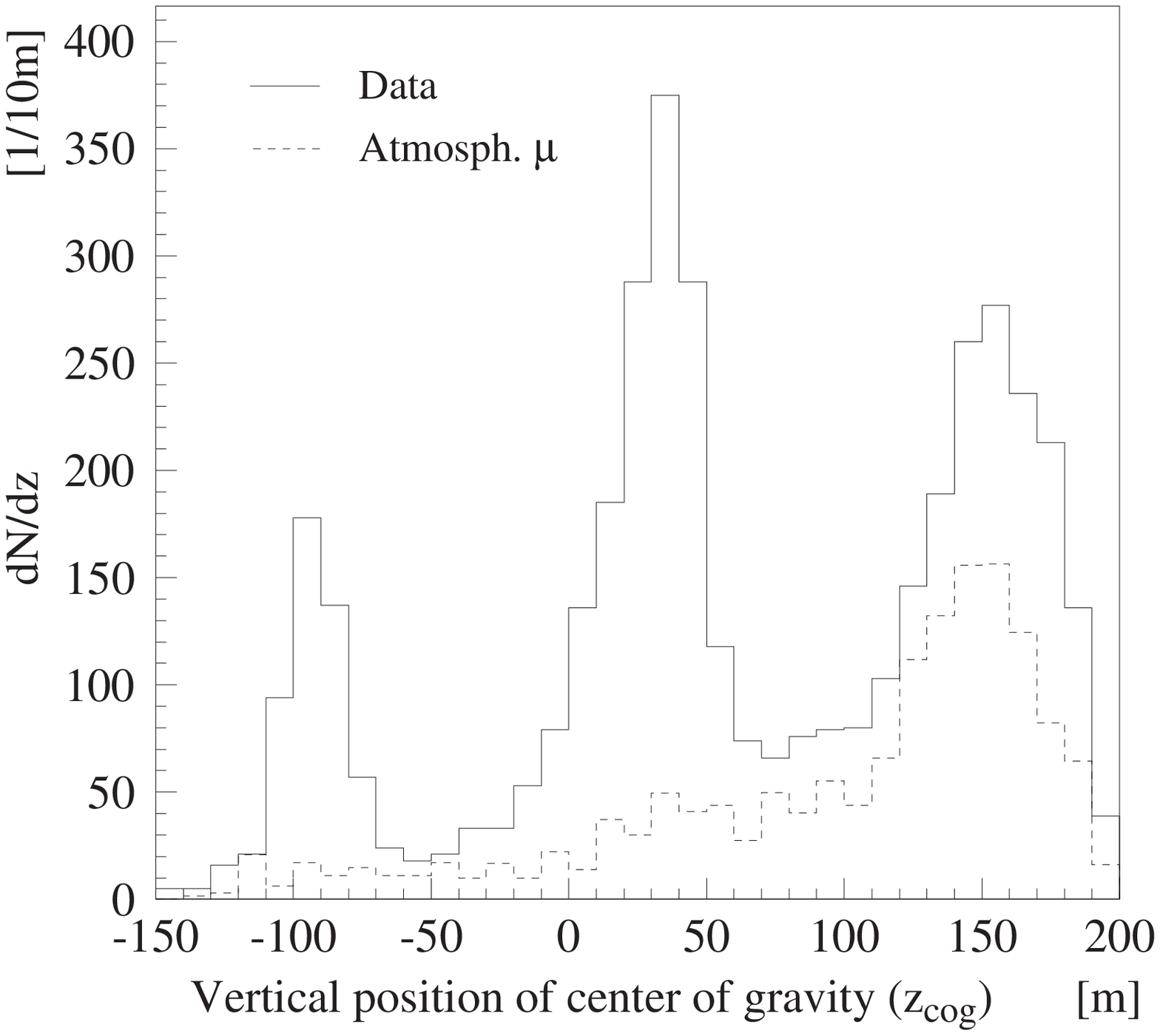} \hfill
  \includegraphics[width=0.85\linewidth]{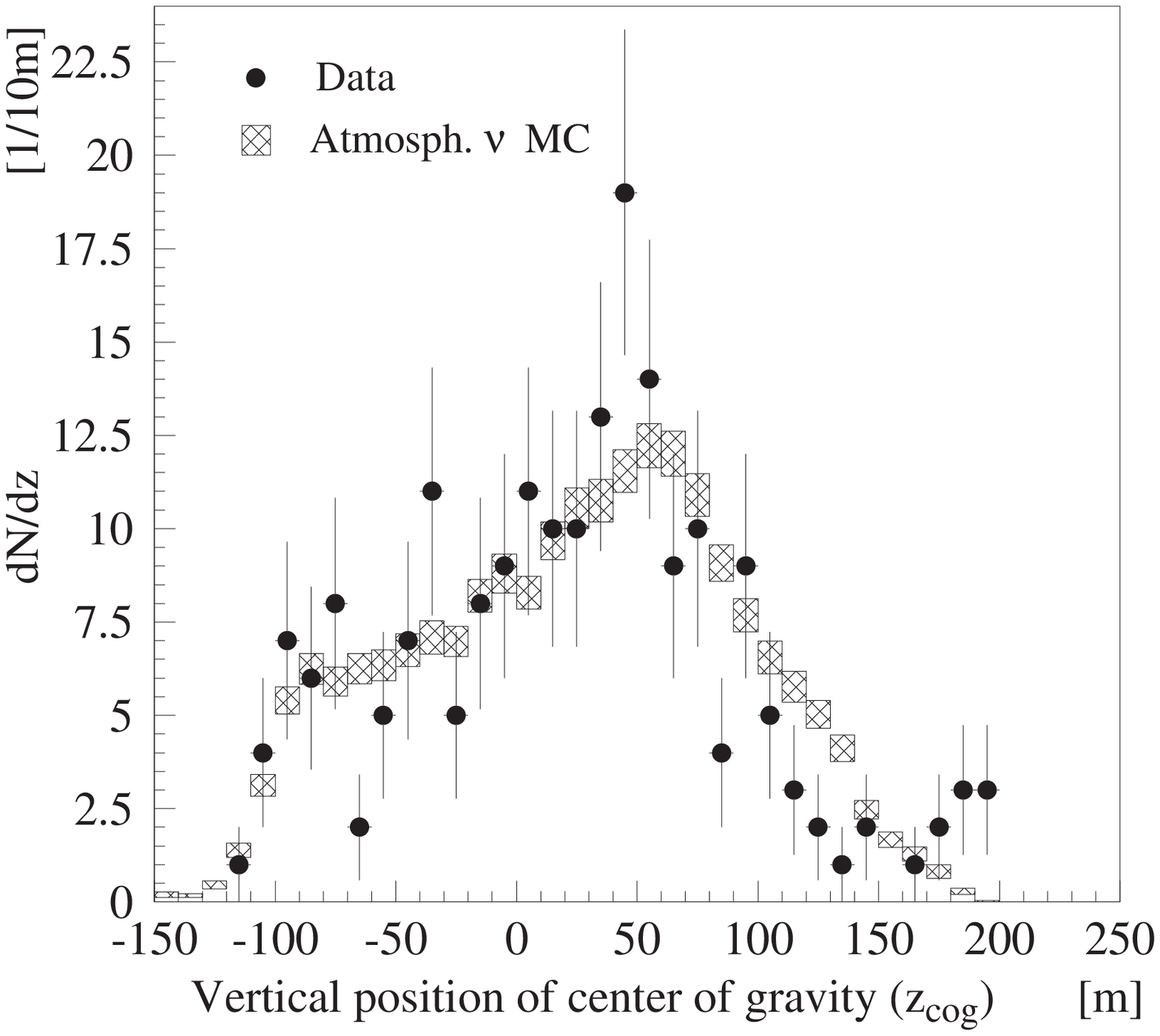}
\caption{Top: Event center of gravity distribution after reconstruction
with special cross-talk cleaning algorithms applied to the events.  
Unsimulated background remains.  Bottom: The data agree with the neutrino
signal after application of additional quality cuts.}
\label{fig:cogzfinal}%
\end{figure}

Other hit cleaning algorithms use the time correlation and amplitude
relationship between real and cross-talk pulses and a map of channels
susceptible to cross-talk and the channels to which they are coupled.  An
additional instrumental effect, believed to be caused by fluctuating high
voltage levels, produces triggers with signals from most OMs on the outer
strings but none on the inner four strings; some 500 of these bogus
triggers were also removed from the data set.  The $5.8 \cdot 10^4$ upgoing
events were again reconstructed after the additional hit cleaning was
applied.  Only $4.9 \cdot 10^3$ (8.4\%) of the events remained upgoing,
compared to 
an expectation from Monte Carlo of 1855 atmospheric muon events (37.8\% of
the total before the additional cleaning), and 555 atmospheric neutrino 
events. 
Figure \ref{fig:cogzfinal} (top)
shows that while there has been a significant reduction in the instrumental 
backgrounds, an unsimulated structure still
remains in the center-of-gravity 
(COG) 
distribution for these remaining data
events. 
The application of additional quality criteria 
brings this distribution in agreement, as shown 
in Figure \ref{fig:cogzfinal} (bottom).

\subsection{Quality Cuts}

The improvements in the reliability of the reconstruction algorithm
described above obviated the need for large numbers of cut parameters or
for careful optimization of the cuts.  Because the signal-to-noise of the
upward-reconstructed data is quite high to begin with, we have the
possibility of comparing the behavior of real and simulated data over a
wide range of cut strengths to verify that the data agree with the
predictions for upgoing neutrino-induced muons, not only in number but also
in their characteristics.  Using the cut parameters described in Section
\ref{subsec:cuts} (with the likelihood replaced by the Bayesian posterior
probability) and a requirement that events  
fitted as relatively horizontal by
the line fit filtering algorithm not be reconstructed as steeply upgoing by
the full reconstruction (a requirement that suppresses residual cross-talk
misreconstructions), an index of event quality was formed.

To do so, we rescale the six quality parameters described above by the
cumulative distributions of the simulated atmospheric neutrino signal, and
consider the six-dimensional cut space formed by the rescaled parameters. 
A point in this space corresponds to fixed values of the quality
parameters, and events can be assigned to locations based on their track
length, sphericity, and so forth.

\begin{figure}
  \centering
  \includegraphics[width=.80\linewidth]{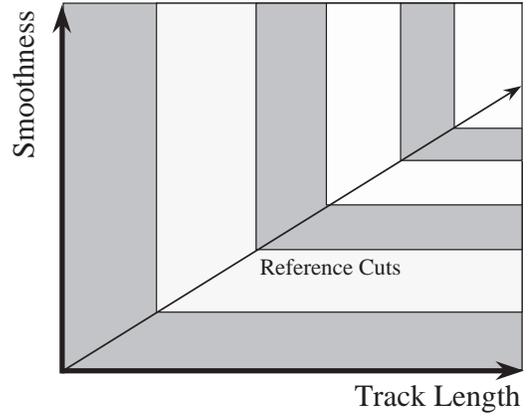}
  \caption{Definition of event quality.  Events are plotted in
  $N$-dimensional cut space (two dimensions are shown here for clarity).  A
  line is drawn from the origin (no cuts) through a selected set of cuts, and
  the space is divided into rectangular shells of equal width.  Events are
  assigned a quality $q$ according to the shell in which they are found.}
  \label{fig:cut_space}
\end{figure}

It is difficult to compare the distributions of data and simulated up- and
downgoing muons directly because of the high dimensionality of the space. 
We therefore project the space down to a single -- ``quality'' --
dimension by dividing it
into concentric rectangular shells, as illustrated in
Fig.~\ref{fig:cut_space}.  The vertex of each shell lies on a line from the
origin through a reference set of cuts which are believed to isolate a
fairly pure set of neutrino events.  Events in the full cut space are
assigned an overall quality value, based on the shell in which they lie.

\begin{figure}[ht]
\centering
  \mbox{\epsfig{file=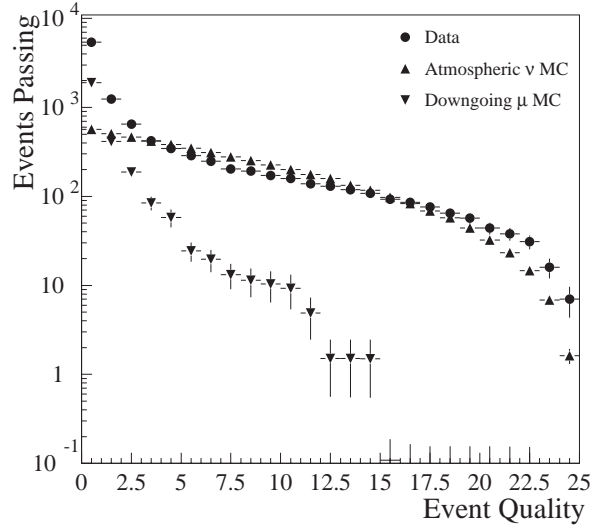,width=7.8cm}}
\caption{
Numbers of events above a certain quality level, for downgoing muon Monte
Carlo, atmospheric neutrino Monte Carlo, and experimental data.  }
\label{fig:qual_a}
\end{figure}

With this formulation we can compare the characteristics of the data to
simulated neutrino and cosmic-ray muon events.  Figure~\ref{fig:qual_a} compares
the number of events passing various levels of cuts; i.e., the integral
number of events above a given quality.  At low qualities, $q\leq3$, 
the data set is dominated by misreconstructed downgoing muons,
data as well as the simulated background exceed the predicted 
neutrino signal.  
At higher qualities, the passing rates of data
closely track the simulated neutrino events, and the predicted background
contamination is very low.

\begin{figure}[ht]
\centering
   \mbox{\epsfig{file=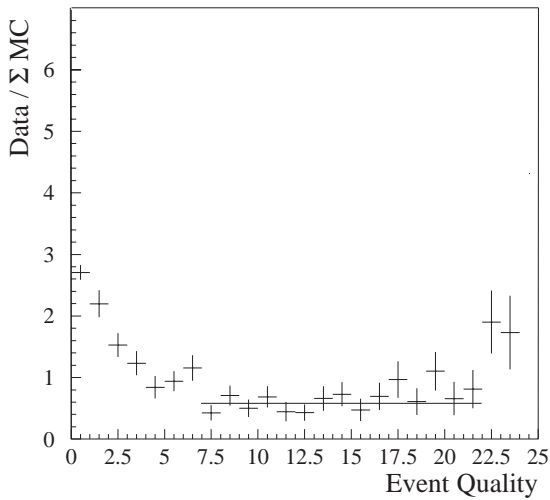,width=7.5cm}}
\caption{Ratio of data to Monte Carlo (cosmic ray muons plus atmospheric
neutrinos).  Unlike Fig.~\ref{fig:qual_a}, the plot is differential --- the
ratio at a particular quality
does not include events at higher or lower
qualities.}
\label{fig:qual_b}
\end{figure}

We can investigate the agreement between data and Monte Carlo more
systematically by comparing the differential number of events \emph{within}
individual shells, rather than the total number of events passing various
levels of cuts.  This is done in Fig.~\ref{fig:qual_b}, where the ratios of
the number of events observed to those predicted from the combined signal
and background simulations are shown.  One can see that at low quality
levels there is an excess in the number of misreconstructed events
observed.  This is mainly due to remaining 
cross-talk.
There is also an excess, though statistically less significant, at very
high quality levels, which is believed to be caused by slight inaccuracies in the
description of the optical parameters of the ice.  Nevertheless, over the
bulk of the range there is close agreement between the data and the
simulation, apart from an overall normalization factor of 0.58. The 
absolute agreement is consistent with the systematic uncertainties. It
should be emphasized that the quality parameter is a convolution of all six
quality parameters, and so the flat line in Fig.~\ref{fig:qual_b}
demonstrates agreement in the correlations between cut parameters.

\begin{table*}[\taboptions]
\newcommand{\m}{\hphantom{$-$}}
\newcommand{\cc}[1]{\multicolumn{1}{c}{#1}}
\renewcommand{\tabcolsep}{1pc} 
\renewcommand{\arraystretch}{1.1} 
\begin{center}
\begin{tabular}{|l|c|c|c|c|}
\hline
          & Monte Carlo     & Monte Carlo &  Data \\       
          & Downgoing $\mu$ & Atmospheric $\nu$  &       \\ 
\hline
Events triggered & $8.8 \cdot 10^8$ & 8978 & $1.05\cdot 10^9$  \\ 
\hline
Efficiency: Reconstructed  upgoing  & $0.55\cdot10^{-5}$ &  & $0.55\cdot10^{-4}$  \\
\hline
 \raisebox{0.5ex}[0pt]{Efficiency: Reconstructed  upgoing}  & 
              \raisebox{-1.5ex}[12pt]{$(2.1\pm0.08)\cdot10^{-6}$} & 
              \raisebox{-1.5ex}[12pt]{$(6.2\pm0.06)\cdot10^{-2}$}& 
              \raisebox{-1.5ex}[12pt]{$4.7\cdot10^{-6}$} \\ 
 \raisebox{0.5ex}[6pt]{(w/ cross-talk cleaning)} & & & \\
\hline
Efficiency: Final Cuts (q$\geq$7) & 
$(1.9\pm0.6)\cdot10^{-8}$ & 
 $(3.1\pm0.03)\cdot10^{-2}$ & $1.9\cdot10^{-7}$  \\
\hline
No. of events: Quality $\geq$ 7 & 17 $\pm$ 5 & 279 $\pm$ 3  & 204\\
\hline
\end{tabular}
\caption{Event numbers for experimental data and Monte Carlo simulations
for four major stages in the analysis.  The errors quoted are statistical
only.  }
\label{table1}
\end{center}
\end{table*}

\subsection{Background Estimation and Signal Description}

\begin{figure}[ht]
\centering
\epsfig{file=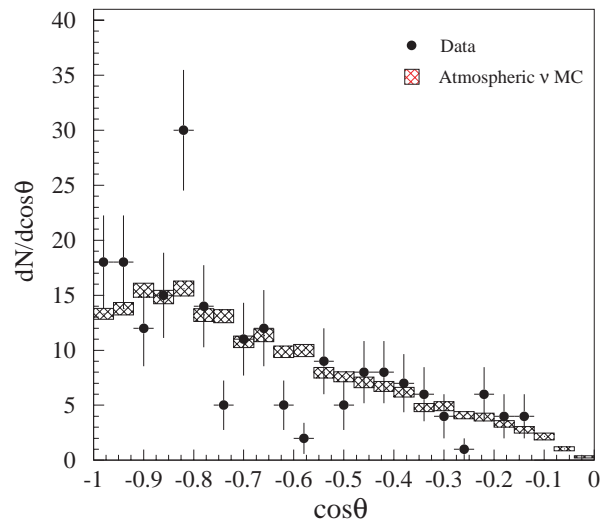,height=7.cm}
\caption{
The zenith angle distribution of upward reconstructed events.  The size of
the hatched boxes indicates the statistical precision of the atmospheric
neutrino simulation. The Monte Carlo prediction is normalized to the   
data. }
\label{fig:zenith}
\end{figure}


If we reduce the 4,917 upward-reconstructed events by requiring a quality
of at least 7 on the scale of Fig.~\ref{fig:qual_a}, we obtain a set of 204
neutrino candidates.  The background contamination, which is due to
misreconstructed downgoing muons, was estimated in three ways.  The
first way is to simulate the downgoing muon flux,
bearing in mind that we are looking at a very low tail ($10^{-8}$) of 
the total
muon distribution.  The second way is to renormalize the signal
simulation by the factor of 0.58 obtained from Fig.~\ref{fig:qual_b} and
subtract the predicted events from the observed data set (accepting the
excess at extremely high qualities, however, as signal).  The third
way, a cross check on the first two methods, is to examine the
data looking for fakes due to unsimulated effects such as cross-talk,
independent coincident downgoing muons, and so forth.  All three methods
yield estimates of 5--10\% contamination.

\begin{figure}[ht]
  \centering
  \includegraphics[width=0.86\linewidth]{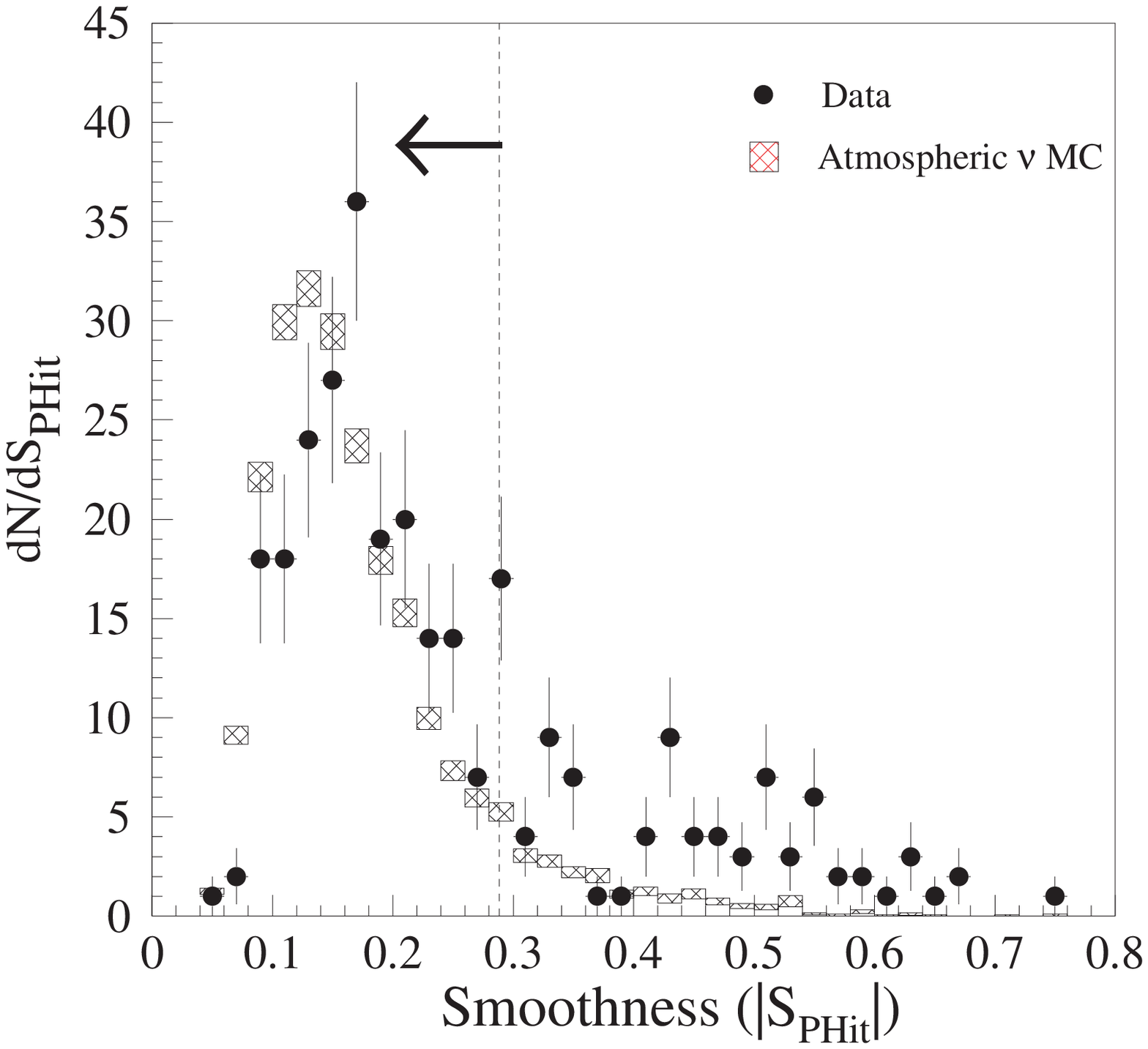} \hfill
  \includegraphics[width=0.86\linewidth]{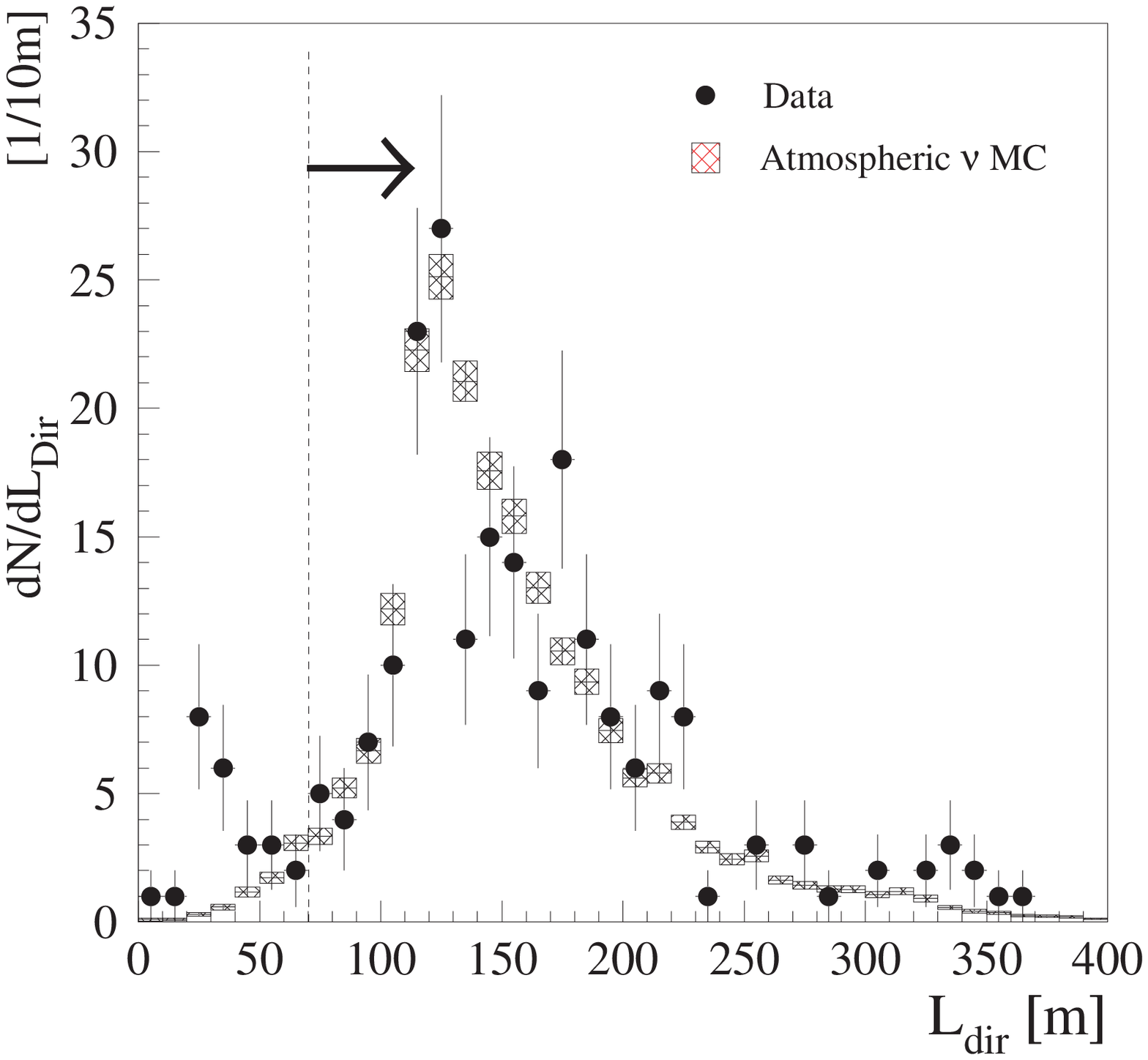}
\caption{Smoothness and direct length variables where quality level 7
cuts have been applied in all but the displayed variable
($N-1$ cuts, see also Section~\ref{sec:nu_character}, Fig.~11).  
The vertical
dashed lines with the arrow indicate the region of acceptance in the 
displayed variable. 
In each case, a clear tail of fake events is removed by application of the
cut, leaving good agreement in shape between the remaining events and
the Monte Carlo expectation.}
\label{fig:nminus1plots}
\end{figure}

The zenith angle distribution for the 204 events is shown in Figure~\ref{fig:zenith}, 
and compared to that for the simulation of atmospheric
neutrinos.  In the figure the Monte Carlo events are normalized to the
number of observed events to facilitate comparison of the shapes of the
distributions.  The agreement in absolute number is consistent with the
systematic uncertainties in
the absolute sensitivity and the flux of high
energy atmospheric neutrinos.  The shape of the distribution of data is
statistically consistent with the prediction from atmospheric neutrinos. 
Figure~\ref{fig:cogzfinal} (bottom) shows the distribution of 
the $z_{\mathrm{COG}}$ parameter for
the 204 events.  The level 7 quality cuts have removed the remainder of the
instrumental events left after the Bayesian reconstruction with the
improved cross-talk cleaning algorithm,
bringing the data events in line with the atmospheric neutrino
expectations.
The efficiencies corresponding to the three steps of the 
data analysis: 1) events reconstructed upward, 2) events reconstructed upward 
with cross-talk cleaning, and 3) with additional level 7 quality cuts 
are summarized in table \ref{table1}.

Figure~\ref{fig:nminus1plots} (top) shows the smoothness distribution for events
that have passed the quality level 7 cuts for the five observables except
smoothness.  The vertical dashed line at smoothness $\sim 0.29$ shows the
value of the level 7 smoothness cut.  This cut removes the tail of fake
events leaving a good agreement between remaining data and Monte Carlo. 
Figure~\ref{fig:nminus1plots} (bottom) shows the same plot for the direct length
variable.  Again, a clear tail of fake events is removed by requiring a
direct length of greater than 70 meters.


\section{Systematic Uncertainties \label{sec:syst}}

As a novel instrument, AMANDA poses a unique challenge of calibration. 
There are no known natural sources of high energy neutrinos, apart from 
atmospheric neutrinos, whose
observation could be used to measure the detector's response.  Understanding
the behavior of the detector is thus a difficult task, dependent partly on
laboratory measurements of the individual components, partly on
observations of artificial light sources embedded in the ice, and partly on
observations of downgoing muons.  Even with these measurements,
uncertainties in various properties that systematically affect the response
of the detector persist, which prevent us at this time from making a
precise measurement of the atmospheric neutrino flux.  The primary
sources of systematic uncertainties, 
and their approximate effects on the number of
upgoing atmospheric neutrinos in the final data sample, as determined by
variation of the simulations, are listed below.

As discusssed in Sections 2 and 3, 
AMANDA is embedded in a natural medium, which is the result of millennia of
climatological history, that has left its mark in the form of
layers of particulate matter affecting the optical properties of the
ice.  Furthermore, the deployment of optical modules requires the melting
and refreezing of columns of the ice.  This cycle results in the formation
of bubbles in the vicinity of the modules, which increase scattering and
affect the sensitivity of the optical modules in ways that 
are not yet fully understood.  The effects of this local hole ice 
are difficult to
separate from the intrinsic sensitivity of the OMs.  The uncertainties in
the neutrino rate are approximately 15\% from the bulk ice layer modeling
in the Monte Carlo, and as much as 50\% from the combined effects of the
properties of the refrozen 
hole ice close to the OMs, and the  
intrinsic OM sensitivity, and angular response.

\begin{figure}[ht]
  \centering 
  \hskip -0.0cm
  \includegraphics[width=0.9\linewidth]{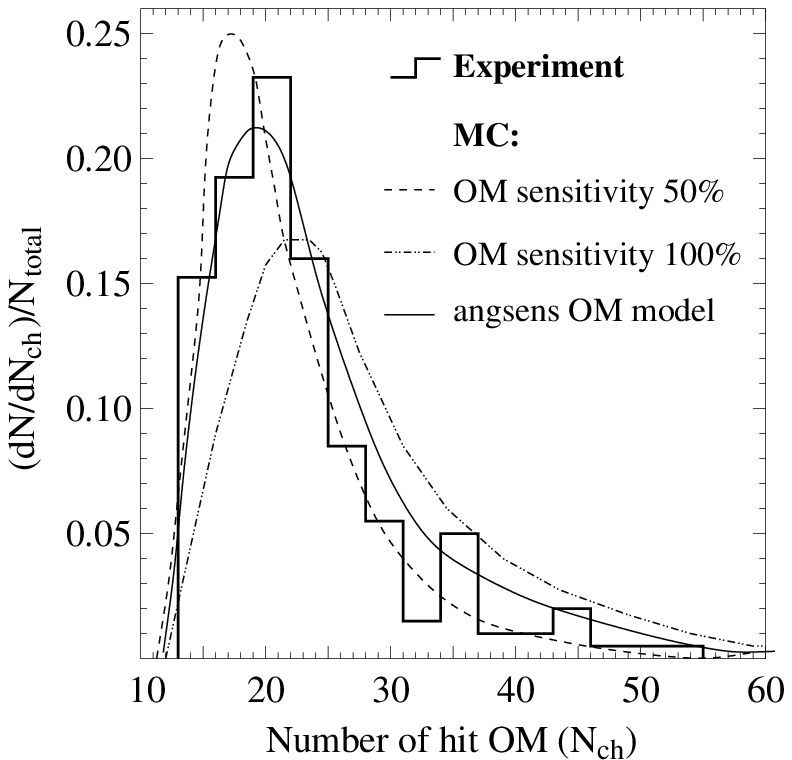}
\vskip 0.3cm
  \includegraphics[width=0.90\linewidth]{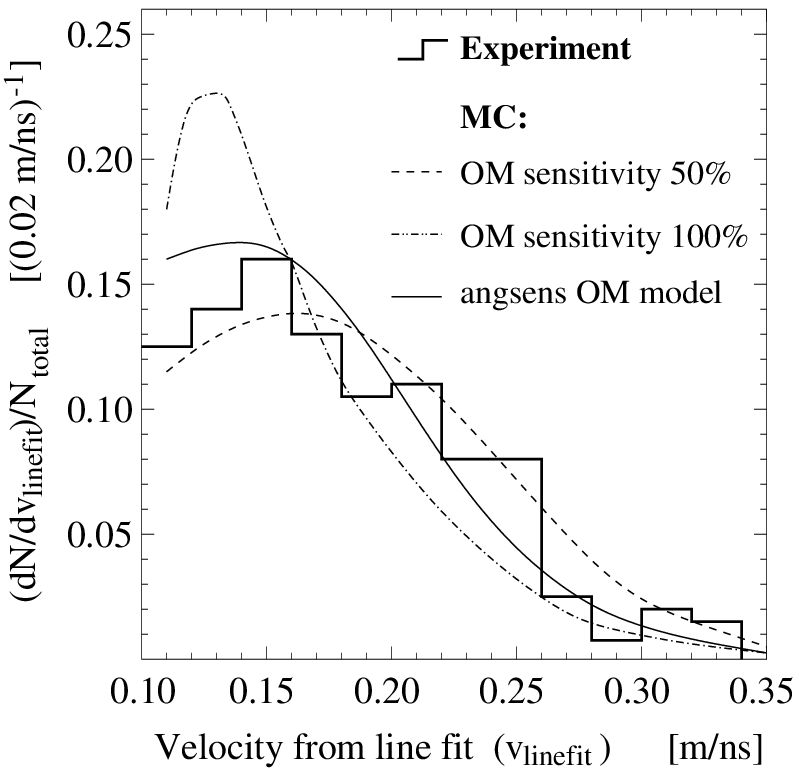}
\caption{
Distributions of two 
variables that are affected by the OM sensitivity, comparing different
signal Monte Carlos to the observed data.  Top:
the number of OMs hit ($N_{\mathrm{ch}}$); 
bottom:
the event velocity for a simplified fit (line fit, $v_{\mathrm{linefit}}$). }
\label{fig:linefit}
\end{figure}

Figure~\ref{fig:linefit} shows two 
variables that
are sensitive to the absolute OM sensitivity: the number of 
OMs hit
and the velocity of the line fit.  
The systematic effects of varying OM sensitivity on the hit multiplicity for
Analysis~I are shown on the top.
The peak of the multiplicity distribution for the standard Monte Carlo
(nominal efficiency 100\% --- dashed line) 
lies at a higher value than for the data.  Reducing the simulated OM sensitivity 
by 50\% results in a peak at lower
values than the data. 
The other variable strongly affected by the OM sensitivity -- 
the velocity of the line fit, introduced in Section \ref{sec:clean_filter}) -- 
is the apparent velocity of the observed light front 
traveling through the ice, see Fig.~\ref{fig:linefit} (bottom).

As a next step, we investigated the effect of the 
 {\tt angsens} OM model (first introduced in Section 3)
on the atmospheric neutrino Monte Carlo simulation.
The results of this simulation gave a more consistent
description of the experiment for several variables --- e.g., the hit
multiplicity (the dotted line in Fig.~\ref{fig:linefit}) --- and they 
produced the absolute neutrino event prediction closer to what was found
in Analysis~I (236.9 events
predicted, 223 observed).  Similar effects are seen when this Monte Carlo
is used with Analysis~II, however the number of predicted events is 25\%
smaller than observed.  Thus the {\tt angsens} model, while
encouraging, does not completely predict the properties of observed events
in both analyses.

Another uncertainty lies in the Monte Carlo routines used to propagate muons
through the ice and rock surrounding the detector.  A comparison of codes
based on
\cite{lohmann} and~\cite{lipari} indicates that different propagators may
change the event rates by some 25\%.

\begin{figure}
\hskip -0.5cm
\includegraphics[width=0.9\linewidth]{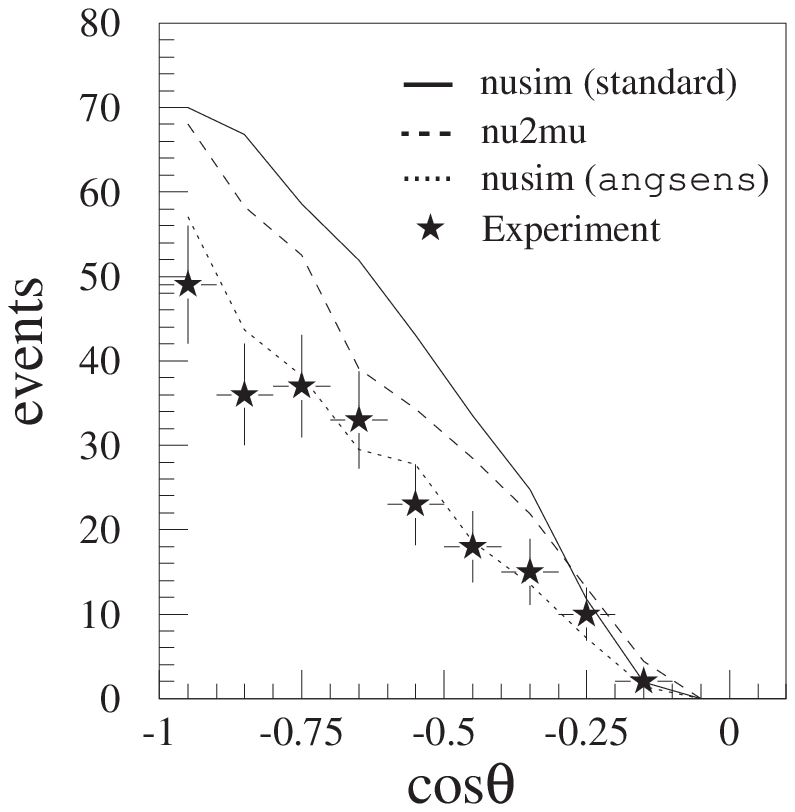}
\caption{Distribution of simulated zenith angles 
for different neutrino generators and also for
a modified angular sensitivity of the OM.
}
\label{fig:sys_zenith}
\end{figure}

Other factors include the simulation of the data acquisition electronics
and possible errors in the time calibrations of individual modules.  These
effects have been studied by systematically varying relevant parameters in the
Monte Carlo simulations.  
For realistic levels of variation, these effects are
well below the 10\% level.

Figure~\ref{fig:sys_zenith} demonstrates how the zenith angle distribution
depends on different atmospheric neutrino event generators (our standard
generator {\tt nusim}~\cite{nusim} and another generator 
{\tt nu2mu}~\cite{Eva}), and also on the
chosen angular sensitivity of the optical module.  
Neutrino flavor oscillations  lead
to a further reduction of the {\tt angsens} prediction by 5.4\% (in
particular, close to the vertical direction), 
assuming $\sin^2 2\theta=1$ and $\Delta m^2=2.5\cdot10^{-3}\;
\mathrm{eV}^2$ \cite{SK2}. The prediction is reduced 
by 11\% if the largest allowed  $\Delta m^2$ is used.

\begin{figure}
\centering
\includegraphics[width=0.95\linewidth]{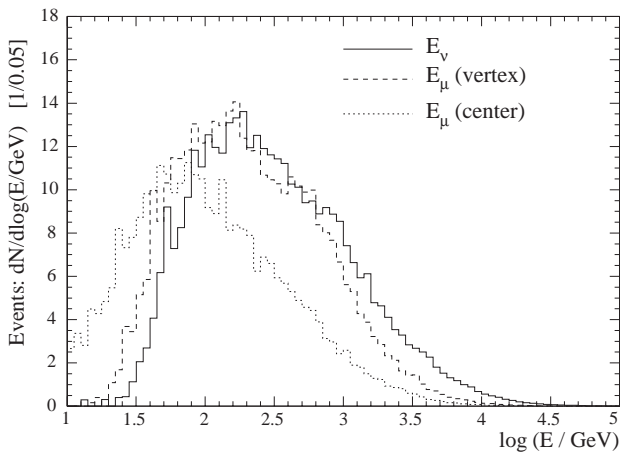}
\caption{Energy distributions for simulated atmospheric neutrino events
which pass the final neutrino cuts.  The effect of neutrino 
oscillations has not been
taken into account.  The figure shows the neutrino and muon energies at the
interaction vertex and the energy of the muons at the point of closest
approach to the detector center.  }
\label{fig:eneff}
\end{figure}

The combined effect of all these systematic uncertainties 
is sufficiently large that
simulations of a given atmospheric neutrino flux can produce predictions
for the event rate varying by a factor of two.
By contrast, the estimated theoretical uncertainty in the atmospheric
neutrino flux, at the energies probed by these analyses, is 
30\%~\cite{gaisser}.  The
effect of neutrino oscillations with the Super-K preferred parameters would
be less than 10\% at these energies.

\section{Synthesis and General Overview \label{sec:conclusion}}

Both Analyses~I and~II are able to separate more than 200 neutrino event
candidates from the 130.1 days of AMANDA-B10 detector livetime in 1997. 
Based on atmospheric neutrino simulations we find that
about 4\% of the total number of events triggered by
upward moving neutrinos passed the final selection.  
A total deficit in the event rate of about 35\% with respect
to the standard neutrino Monte Carlo prediction is found for both analyses.
An event overlap of 102 experimental events is observed, consistent with a
predicted overlap of $119 \pm 13$ from the atmospheric neutrino Monte Carlo. 
Thus, the combined sample of data provides about 300 neutrino candidates.
Both analyses estimate their residual background to be 
about 10\% of the number of neutrino event candidates.

\begin{figure}
\centering
\includegraphics[width=0.95\linewidth]{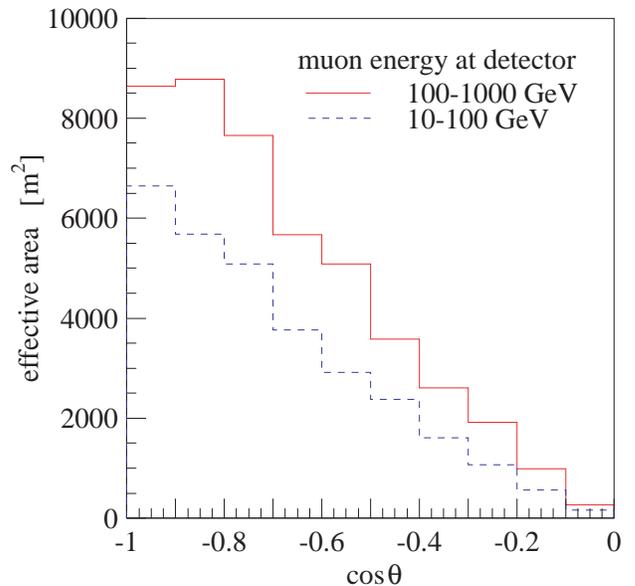}
\caption{Effective area for muons versus 
zenith angle. The energy of the muons is
given at the point of closest approach to the detector. }
\label{fig:effarea}
\end{figure}

Figure~\ref{fig:eneff} shows the energy distribution of the simulated
neutrinos and the corresponding muon events.  Ninety percent of all Monte
Carlo signal events have muon (neutrino) energies between 48 (66) GeV and
1.8~(3.4)~TeV\@.  The dominant part of the signal events in this analysis
comes from neutrino energies below 1~TeV.
Figure~\ref{fig:effarea} shows the
effective area as a function of the zenith angle for two ranges of the muon
energy at the point of closest (POC) approach to the detector.  The effective
area for muons with energies at POC between 100 and 1000\,GeV 
is $ 3.9\cdot10^4\,$m$^2$ at trigger level and 
 $2800\,$m$^2$ after application of the neutrino selection cuts. 
It should be noted that much higher effective
areas are possible when searching for neutrinos from astrophysical point
sources \cite{point_tbs} or from gamma ray bursts \cite{GRB_ICRC}.

\begin{figure}
\centering
\includegraphics[width=0.95\linewidth]{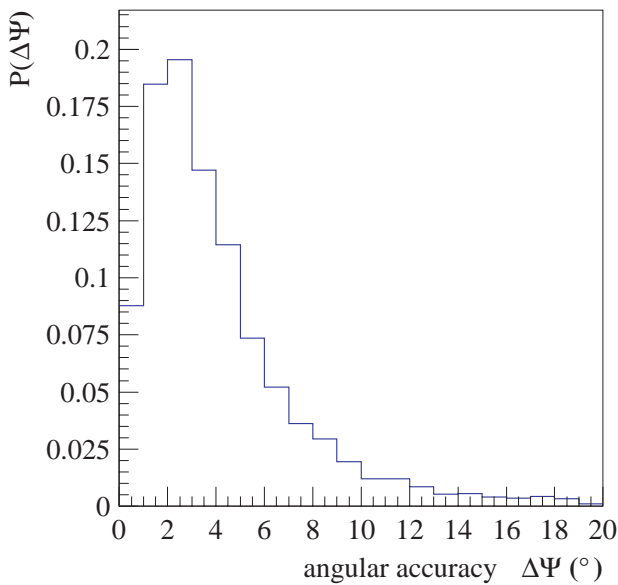}
\caption{Monte Carlo simulation of the angular resolution 
for muons that pass the final selection 
criteria. The median error is $3.2^\circ$.}
\label{fig:angres}
\end{figure}

\begin{figure}[ht]
\centering
\includegraphics[width=1.0\linewidth]{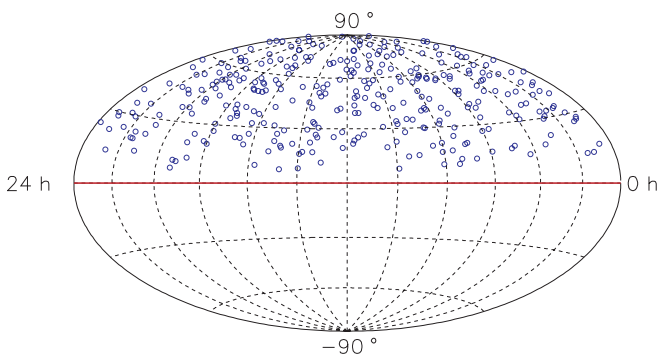}
\caption{Neutrino 
skyplot of upgoing events as seen with AMANDA-B10 in 1997 
in equatorial coordinates. 
In this
figure, neutrino events from both analyses are combined.  The background of
non-neutrino events is estimated to 
be less than 10$\%$.  }
\label{fig:sky}
\end{figure}

Figure~\ref{fig:angres} shows the point spread function 
of the reconstructed muon trajectory with 
respect to the true muon direction. 
Based on Monte Carlo simulations, we find a median angular resolution 
of muons from atmospheric neutrinos of $3.2^\circ$
for the final sample. 
A more detailed study of the angular resolution
can be found in \cite{BIRON, spase_icrc, spase_calib}. 
Figure~\ref{fig:sky} shows the skyplot (equatorial coordinates) 
of all the candidate neutrino 
events found across both analyses. The distribution 
of the events on the skyplot is consistent with 
a random distribution.

\section{Conclusions}
The AMANDA-B10 data from 130.1 days of livetime during the austral winter
of 1997 have been analyzed in an effort to detect high energy atmospheric 
neutrino events, and to compare their properties to expectations.  
Two working groups
in the collaboration, using differing reconstruction, cut optimization and
instrumental event rejection techniques, produced sets of 223 and 204
neutrino candidates, respectively.  
Several methods of background estimation put the
residual event contamination from downgoing atmospheric muons and 
instrumental artifacts at about 10\%.  
Taking into account systematic uncertainties, the observed event numbers are
consistent with 
systematically varied atmospheric neutrino Monte Carlo
predictions, which are from 150--400 events.  The range of these
predictions is dominated by uncertainties in the neutrino flux, in the
understanding of photon propagation through the bulk ice and the
refrozen hole ice, and in muon propagation and energy loss.  
The Monte Carlo suggests that
 90\% of the selected events are produced by neutrinos in the energy
range of $\sim 66$\,GeV to $3.4$\,TeV\@.  The observation of
atmospheric neutrinos in line with expectations establishes AMANDA-B10 as a
working neutrino telescope.  We finally note that many of the procedures
for signal separation simplify considerably in larger detectors.  In
particular, first results from AMANDA-II \cite{ama_ii_1} demonstrate that
the neutrino signal is separated with much higher efficiency and with fewer
cuts than for AMANDA-B10.

\paragraph*{Acknowledgments}
This research was supported by the following agencies: U.S.
National
Science Foundation, Office of Polar Programs; U.S. National
Science
Foundation, Physics Division; University of Wisconsin Alumni
Research
Foundation; U.S. Department of Energy; Swedish Natural
Science Research
Council; Swedish Research Council; Swedish Polar Research Secretariat; Knut and Alice
Wallenberg
Foundation, Sweden; German Ministry for Education and
Research; U.S.
National Energy Research Scientific Computing Center
(supported by the
Office of Energy Research of the U.S.
Department of Energy); UC-Irvine AENEAS Supercomputer
Facility; Deutsche
Forschungsgemeinschaft (DFG). D.F. Cowen acknowledges the
support of the
NSF CAREER program and C. P\'erez de los Heros acknowledges
support from
the EU 4th framework of Training and Mobility of
Researchers.


\end{document}